\PassOptionsToPackage{dvipsnames}{xcolor}
\documentclass[sn-mathphys,Numbered]{sn-jnl}

\usepackage{graphicx}\usepackage{multirow}\usepackage{amsmath,amssymb,amsfonts}\usepackage{amsthm}\usepackage{mathrsfs}\usepackage[title]{appendix}\usepackage{xcolor}\usepackage{textcomp}\usepackage{manyfoot}\usepackage{booktabs}\usepackage{algorithm}\usepackage{algorithmicx}\usepackage{algpseudocode}\usepackage{listings}

\usepackage{amsmath}
\usepackage{bbm}
\usepackage{dsfont}
\usepackage{bm}
\usepackage{xspace}
\usepackage{hyperref}
\usepackage[nameinlink,capitalize]{cleveref} 

\usepackage{subcaption}
\usepackage[utf8]{inputenc}
\usepackage{microtype}
\usepackage[htt]{hyphenat}
\usepackage{tensor}
\usepackage{tabularx}
\usepackage{paralist}
\usepackage{xcolor}
\usepackage{booktabs}
\usepackage{amsfonts}
\usepackage{amsthm}
\usepackage{epigraph} 
\usepackage{acronym}

\DeclareMathOperator{\His}{\mathcal{H}}

\newcommand{\titlename}{Conductance and Influence-Capital: Modeling Online Social Influence}

\newcommand{\gim}{Generalized Influence Model\xspace}
\renewcommand{\gim}{GIM\xspace}

\newcommand{\covid}{\textsc{\#Covid-19}\xspace}

\newcommand{\Hawkesinf}{Hawkes-modeled Influence\xspace}
\newcommand{\hawkesinf}{Hawkes-modeled influence\xspace}
\newcommand{\socialcapital}{influence-capital\xspace}
\newcommand{\Socialcapital}{Influence-capital\xspace}
\newcommand{\ndcgauc}{NDCG-AUC\xspace}

\newcommand{\sh}[1]{\textbf{#1}}

\definecolor{navy}{rgb}{0.1, 0.1, 0.8}
\definecolor{gray}{rgb}{0.4, 0.4, 0.4}
\definecolor{olive}{rgb}{0.1, 0.5, 0.1}
\definecolor{ruby}{rgb}{0.8, 0.1, 0.3}
\definecolor{darkpastelgreen}{rgb}{0.01, 0.75, 0.24}
\definecolor{celestialblue}{rgb}{0.29, 0.59, 0.82}
\definecolor{coral}{rgb}{1.0, 0.5, 0.31}
\definecolor{blue}{rgb}{0.23, 0.44, 0.62}
\definecolor{Goldenrod}{rgb}{0.8,0.8,0}

\newcommand{\eat}[1]{}

\newcommand{\revEPJA}[1]{{#1}}
\newcommand{\replaceEPJA}[2]{#1}

\newcommand{\NOTE}[2]{}
\newcommand{\note}[1]{}
\newcommand{\editnote}[2][1=]{}
\newcommand{\nb}[1]{}
\newcommand{\mar}[1]{}
\newcommand{\roh}[1]{}
\newcommand{\TODO}[2]{}

\theoremstyle{thmstyleone}

\theoremstyle{thmstyletwo}

\theoremstyle{thmstylethree}

\begin{document}

\title[\titlename]{\titlename}

\author*[1]{\fnm{Rohit} \sur{Ram}}\email{rohit.ram@student.uts.edu.au}

\author[1]{\fnm{Marian-Andrei} \sur{Rizoiu}}\email{marian-andrei.rizoiu@uts.edu.au}

\affil*[1]{\orgname{University of Technology Sydney}, \country{Australia}}

\abstract{
    Human interactions are mediated by social influence.
    During crises like the COVID-19 pandemic, social influence can be critical in determining whether life-saving information is adopted, public health measures are observed, or immunization campaigns meet their targets. 
    The literature on online social influence presents notable limitations across disciplines. 
    Psychosocial approaches effectively characterize the nature of influence by measuring how social factors impact these phenomena, but they lack computational modeling capabilities and rely on slow, non-scalable measurement methods that struggle to generalize to current issues. Conversely, computational approaches, while data-driven and based on network and event analysis, often fail to incorporate the critical social factors that underlie these phenomena.
    Our work bridges this gap through two main contributions that integrate the strengths of both approaches. 
    First, we present a data-driven Generalized Influence Model that incorporates two novel psychosocial-inspired mechanisms: the conductance of the diffusion network and the \socialcapital distribution. This model not only outperforms existing state-of-the-art approaches but also corrects the inherent biases introduced by the widely used follower count metric. Second, we empirically test long-held sociological hypotheses regarding influence, social class, and expertise in the online domain by applying our influence model to discussions around COVID-19. 
    We quantify the influence and content veracity for more than 21.5 million X/Twitter users in relation to their professions. 
    \replaceEPJA{Our model suggests that executives, media, and military figures exert greater influence than pandemic-related experts such as life scientists and healthcare professionals.}{We discover that executives, media, and military figures exert greater influence than pandemic-related experts such as life scientists and healthcare professionals.}
    Worryingly, by leveraging existing COVID-19 misinformation datasets, we show that some of the most influential occupations also spread the most misinformation. These findings raise questions about the effectiveness of information dissemination by experts in situations of crisis.
}


\maketitle

\section{Introduction}
\label{gim:sec:introduction}
Understanding public opinion formation remains an enigma with ample impact on our increasingly digitized society.
The polarization of public opinion signals breakdowns in social trust \citep{rapp2016moral}, extreme opinions form pathways to radicalization \citep{mccauley2008mechanisms}, and in recent years, opinions opposing expert medical advice have led to loss of life by undermining immunization campaigns and stoking vaccine hesitancy.
While we are beginning to understand the factors that drive opinion formation, we lack scalable modeling and quantification tools.
These tools are essential when trying to mobilize our peers to enact change---whether to accept the vaccine or alter their consumption patterns to avert catastrophic climate change.
One of the foundational factors in opinion formation and in enacting change is \emph{influence} \citep{nowak2005dynamics}.
This force governs interpersonal relationships and contributes to establishing societal institutions and reforms.

Social media ubiquity has expanded the role of influence by providing a fertile ground for influence mechanisms to unfold and for a minority of users to exert disproportionate control.
There exists ample evidence that online dynamics have offline repercussions in terms of collective movement \cite{greijdanus2020psychology} and radicalization \cite{gunton2022impact}.
Although there is a rich trove of literature on social influence, stretching from the psychosocial to computational domains, a quantitative approach to measuring online influence remains elusive~\citep{peng2018influence, mason2007situating}.
The computational approaches focus heavily on measuring message propagation propensities and exposure to messages through one's social network \citep{kempe2003maximizing,du2013scalable,mishra2016feature}; however, social influence is more accurately defined as the ability to change the behaviors, opinions, or beliefs of others \citep{moussaid2013social}.
Many psychosocial complexities of influence (like the social attributes of the source of influence) are lost when influence is narrowly interpreted as merely a reaction to exposure.
Therefore, we ask \textbf{how can online influence be quantified, such that it relates to the theoretical psychosocial influence phenomenon?}

Furthermore, sociologists have long espoused that influence is most directly related to concepts such as social class and expertise \citep{kraus2019social}.
Social class, a multifaceted construct encompassing education, wealth, occupation, and subculture, strongly aligns with established determinants of influence such as authority and likability \cite{cialdini2001science}.
While historically linked to political power, in contemporary egalitarian societies with social mobility, class is predominantly indicated by occupation.
Expertise, on the other hand, is related to a specific context and typically derives influence through demonstrated outcomes.
Despite the compelling appeal of these sociological theories of influence, they remain largely untested on a large scale in online domains.
Evaluating these theories is crucial, as adherence to expert advice reflects optimal societal behavior, and the extent to which social class either hinders or reinforces such advice can have critical implications.
For instance, in an ideal scenario, populations would follow expert advice (e.g., from epidemiologists, biomedical scientists, healthcare professionals) during pandemics.
However, the emergence of the anti-vaxxer movement and the proliferation of vaccine misinformation illustrates a contrasting reality.
Therefore, we ask \textbf{how does influence correlate with its proposed determinants in the online domain?}

\begin{figure}[tbp]
    \includegraphics[width = 0.9\columnwidth]{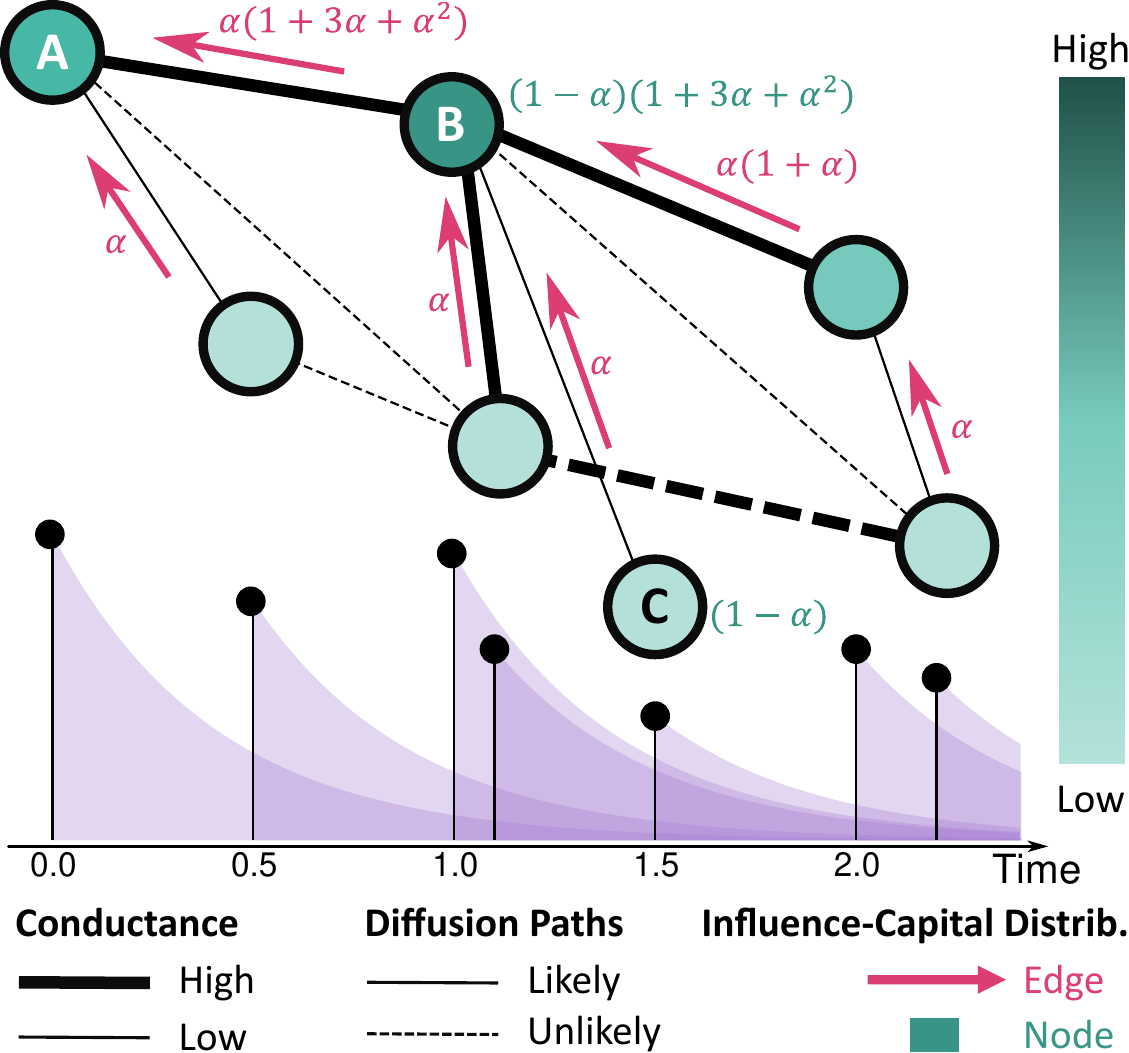}\caption{
        Schema of the Generalized Influence Model (GIM).
        (bottom) An example cascade is modeled using Hawkes processes.
        Each event (timestamp on x-axis) has a mark (y-axis) and spawns new events following a time-decaying intensity (\textcolor{Purple}{magenta} areas).
        (top) The latent branching structure is shown with solid lines, and other valid pathways are shown with dotted lines.
        \gim has two psychosocial-inspired components.
        \emph{Conductance}: Edge thickness represents conductance, which modulates the likelihood of observing diffusions along that edge.
        \emph{\Socialcapital distribution}: A percentage $\alpha$ of a node's capital (\textcolor{PineGreen}{green} shades) is transferred along diffusion edges (\textcolor{RedViolet}{red} arrows), from target to source.
        Influence is proportional to the accumulated capital.
    }
    \label{gim:fig:teaser}
\end{figure}

In this paper, we model online influence on Twitter/X.
We achieve this in a sequence of two parts.
In \emph{the first part}, we build the Generalized Influence Model (\gim) -- a novel contribution that bridges the divide between psychology and the quantitative approaches~\citep{asch1961effects}.
In \cref{gim:sec:preliminary}, we review an essentially quantitative model \cite{rizoiu2018debatenight} for online influence based on exposure.
In \cref{gim:sec:gim}, we build \gim by augmenting this model with two factors known in psychosocial literature to modulate the exertion of social influence: \emph{conductance} \cite{cialdini2001science, newcomb1953approach} and \emph{social attribution} \cite{milgram1978obedience, horai1974effects,cialdini2001science,ghaffar2020structural}.
We endogenize these factors by incorporating a social network conductance and a \socialcapital distribution mechanism and propose several flexible implementations that maintain scalability.
In \cref{gim:subsec:gim-evaluation}, we show \gim to outperform several baselines, including the current state-of-the-art influence quantification~\citep{rizoiu2018debatenight}, and to correct the biases introduced by the widely used follower count metric.
In \emph{the second part}, in \cref{gim:subsec:occupational-application}, we empirically test the hypothesis that social class and expertise are the primary influence determinants.
Following prior work, we assume that social class is primarily associated with one's occupation \cite{sloan2015tweets} and expertise is associated with the quality of the sources they share.
On a Twitter/X dataset containing discussions about the COVID-19 pandemic, we utilize \gim to identify the occupational groups that yield the highest influence.
We determine user occupation using the O*NET taxonomy~\cite{onet}, and numerically quantify the veracity of information users spread using a dataset of COVID-19 related misinformation~\citep{cui2020coaid}.
The analysis reveals experts (i.e., epidemiologists, biomedical scientists, and healthcare professionals) are not always the ones that shape the discussions.
So who does exert the greatest influence?

Successfully modeling online influence identifies the actors who shape societal views.
It serves as the first step toward addressing societal issues---such as spreading misinformation amid a pandemic.
\textbf{The main contributions of this work include:}
\begin{compactitem}
    \item The \textbf{General Influence Model} which introduces two psychosocial inspired mechanisms: \emph{conductance} and \emph{\socialcapital distribution}, to move toward psychosocial influence.
    \item An \textbf{evaluation analysis} of GIM, where we find the optimal hyperparameters, show that GIM outperforms the state-of-the-art, and illustrate how GIM is an unbiased estimator.
    \item An \textbf{analysis of sociological determinants of influence}; investigating the relationship between social class, expertise, and GIM social influence.
\end{compactitem}

\revEPJA{To aid the reader, we briefly define the key terms used throughout this paper.
\emph{Conductance} quantifies how readily influence flows between two users: analogous to physical materials conducting electricity or heat, social relationships conduct influence to varying degrees, and higher conductance makes influence transmission more likely.
\emph{\Socialcapital} captures the latent capacity of an individual to exert influence, accumulated through the endorsements they receive in information cascades.
The \emph{branching structure} is the latent tree of who-influenced-whom within a retweet cascade.
\emph{Ground truth influence} refers to a peer-perceived influence ranking obtained via crowdsourced pairwise comparisons, used to calibrate and evaluate our model.}

\section{Preliminaries \& Related Works}
\label{gim:sec:preliminary}
Quantitative influence approaches exploit readily available metadata, such as timing and network features, which has historically been underutilized in psychosocial literature. 
These approaches, and the axioms from which they are built, form a strong foundation for an inquiry into influence. 
Influence is traditionally difficult to observe, as we cannot easily observe the minds of the influenced.
We build upon prior works that take a longitudinal view and hypothesize that the endorsement patterns over extended periods reveal the influence structures in an observed cohort. 
On Twitter/X, this endorsement is signaled via retweeting, which is widely accepted as a form of endorsement of an emitter (the person being retweeted) by a receiver (the retweeter) \citep{metaxas2015retweets}. However, platforms typically do not expose the structure of these endorsements.

Nearly all social media platforms have a `resharing' behavior (with varying degrees of API accessibility). 
The technique of reconstructing influence structures relies on acquiring the timing of reshare events.

\citet{rizoiu2018debatenight} infer these structures by assuming retweets arrive following a Hawkes point process.
In this section, we review approaches to influence quantification and articulate the Hawkes-modeled influence \cite{rizoiu2018debatenight} from which \gim is built.

\sh{Prior Influence Quantification.}
Common approaches to quantitative models of influence include: using centrality on social graphs as an influence proxy \citep{liu2017influence, silva2013profilerank}, approximating solutions to the influence maximization task \cite{kempe2003maximizing,du2013scalable}, and explicitly modeling popularity through approaches such as Hawkes Processes \citep{mishra2018modeling}. 
However, these approaches suffer from the conflation of social influence with other concepts such as popularity or activity. Several studies have highlighted the discordance between measures of popularity and influence \cite{romero2011influence,bakshy2011everyone,smith2018influence}.
Recent work has proposed several novel approaches to influence measurement.
\citet{nickel2021modeling} efficiently model entity interactions via a Multivariate Hawkes Process (though this approach is entirely based on propagation propensity rather than grounded influence), and \citet{smith2018influence} estimate the impact by removing confounders through a causal inference network framework (though this approach must effectively reconstruct the network, makes restrictive diffusion assumptions, and focuses on influence operator classification). 
Unlike these prior methods, our work directly accounts for psychosocial factors that mediate influence propagation: \socialcapital and network conductance.

\sh{Retweet Cascades}
consist of an original tweet and subsequent retweets. 
We denote a marked cascade up to time $T$ as $\His_\zeta(T)=\{v_1, v_2, \ldots\}$, where $v_i = (t_i, \zeta_i)$ denotes the $i$th tweet, $t_i$ is the event time relative to the original tweet ($t_1 = 0$), and $\zeta_i \in \mathbb{R}$---dubbed the \textit{mark}---is the event meta-data.
We denote a possible realization of the endorsement structure as $\mathcal{G}$, and the set of all potential endorsement structures as $\Upsilon$.

\sh{The Branching Structure} of retweet cascades is a latent graph $(G, E)$ where $G$ are the tweets, and $E$ contains direct retweet relations.
Given a cascade of $n$ events, we define the set of all valid branching structures as $\Upsilon = \{G | (v_i, v_j) \in G \text{ so that } t_i < t_j\}$.
\citet{rizoiu2018debatenight} show that $|\Upsilon| = (n-1)!$.
\cref{gim:fig:teaser} shows an example retweet cascade, its most likely branching structure in solid lines and other potential branching structures in dashed lines.
Next, we compute the probability mass function over the branching structure set $\Upsilon$.

\sh{Hawkes Processes} \citep{hawkes1971spectra} have emerged as a powerful tool for modeling social media events and information diffusion, providing a mathematical framework for capturing the self-exciting nature of online user activities \cite{Rizoiu2017a,Kong2020}.
The self-exciting property---the arrival of an event increases the likelihood of future events--- is applicable here due to the property of social affirmation (i.e., past social actions encourage incoming actions).

In Hawkes processes, events arrive following the conditional intensity
\begin{equation*}
    \lambda(t|\His_\zeta(t)) = \mu(t) + \sum_{v_i \in \His_\zeta(t): t_i < t} \zeta_i^b \phi(t - t_i) \enspace,
\end{equation*}
where $\mu(t)$ is the baseline intensity;
each event $v_i$ increases the overall intensity by $\zeta_i^b \phi(t - t_i)$;
$\zeta_i^b$ is one way to model marks where $b$ mediates the marks' effect, and the kernel $\phi : \mathbb{R}^+ \rightarrow \mathbb{R}^+$ controls the event intensity decay. 
The exponential $\phi_{exp}(t)=e^{-rt}$ and the power-law $\phi_{pow}(t)=(t+c)^{-(1+r)}$ are common parametric forms. 
\looseness=-1 

\citet{hawkes1974cluster} propose the branching representation of Hawkes processes, where each event $v_i$ generates offspring following a non-homogenous Poisson process of intensity $\zeta_i^b \phi(t - t_i)$---illustrated by the magenta areas in \cref{gim:fig:teaser}.
\citet{lewis2011nonparametric} use the branching representation to estimate the probability that $v_j$ is a direct offspring of $v_i$ as 
\begin{equation}
    \label{gim:eq:pij_hawkes}
    p_{ij} = \frac{\zeta_i^b \phi(t_j - t_i)}{\mu(t) + \sum_{t_k < t_j} \zeta_k^b \phi(t_j - t_k)}, t_i < t_j.
\end{equation}
Intuitively, $p_{ij}$ is the proportion of intensity that $v_i$ contributed to the total intensity at time $t_j$. 
Note that for retweet cascades $\mu(t)=0$, and $p_{ij} = 0 , \text{ when } t_i > t_j$.
Finally, the probability of a valid branching structure is
$ \prod_{(v_i, v_j) \in E(G)} p_{ij}$.

\sh{\Hawkesinf.}
\citet{rizoiu2018debatenight} measure influence as the expected number of offspring of a tweet across all valid branching structures. 
They formalize influence as 
$
    \varphi(v_i) = \mathbb{E}_{ \mathcal{G} \in \Upsilon} \Big[ \sum_{t_j > t_i} \mathds{1}(\mathcal{G}_{i \rightarrow j}) \Big] \enspace,
$
where $\mathds{1}(\mathcal{G}_{i \rightarrow j})$ indicates a path between $v_i$ and $v_j$ in the branching structure  $\mathcal{G}$.
Using the \textit{independent cascades} assumption~\citep{kempe2003maximizing} (that generating a tweet at $t_j$ is independent of the diffusion structure up to $t_j$), \citet{rizoiu2018debatenight} devise an efficient iterative procedure for computing $\varphi(v_i)$.
They introduce $m_{ik}$ the pairwise influence exerted by $v_i$ on $v_k$, either directly when $v_k$ is a direct offspring of $v_i$ or indirectly when $v_k$ lies on the same diffusion path as $v_i$.
Formally,
$
    \label{gim:eq:mik}
    m_{ik} = \sum_{j=i}^{k-1} m_{ij}p_{jk} , i<k \enspace,
$
$m_{ik}=1$ when $i=k$, and $m_{ik}=0 \text{ when } i>k$.
Intuitively, $m_{ik}$ is the sum of the probabilities of all valid paths between $i$ and $k$.
Consider an example with three tweets, ${v_1,v_2,v_3}$, the influence $m_{13}$ is computed as
$
    m_{13} = m_{11}p_{13} + m_{12}p_{23} = p_{13} + p_{12}p_{23}
$
representing all possible paths between $v_1$ and $v_3$.
A tweet's influence is the total influence it exerts, i.e. $\varphi(v_i) = \sum_{k=i}^n m_{ik}$. 
The $\{m_{ik}\}$ matrix is computed in $n$ matrix multiplications, and the total time complexity is $O(n^3)$.
Finally, a user's influence is the average influence of all their tweets.

\sh{Towards Psychosocial.}
Our work is inspired by improved performance, via theory-based social features, in tasks related to social influence such as the influence prediction task \citep{luceri2019analyzing,qiu2018deepinf} (classifying whether users perform a behavior after exposure), social graph embedding \cite{gu2018rare,tsitsulin2018verse}, and recommender systems \citep{yu2019spectrum}.
This suggests that incorporating psychosocial theory into computational models \cite{Rizoiu2018} can yield more accurate and nuanced representations of social influence processes.
Our work aims to bridge the gap between computational efficiency and psychosocial theory by introducing mechanisms that capture the relational qualities of influence while maintaining computational scalability.

\section{Methodology}
\label{gim:sec:gim}
Prior work estimates influence through endorsement timing and accessible user metadata; however, these approaches fail to consider the nature of relationships between people and the structure of these relationships.
To address these limitations we propose the Generalized Influence Model (\gim) that quantifies influence based on observed information cascades.
Our proposed \gim generalizes the \hawkesinf (see \cref{gim:sec:preliminary}) by incorporating two mechanisms that model crucial social information, depicted in \cref{gim:fig:teaser}.
These mechanisms -- \emph{conductance} (\cite{cialdini2001science, newcomb1953approach}, see \cref{gim:subsec:conductance}) and \emph{social attribution} (\cite{milgram1978obedience, horai1974effects,cialdini2001science,ghaffar2020structural}, see \cref{gim:subsec:inf-cap-dist})  -- are factors that psychosocial literature has identified as modulating the exertion of social influence.

We first formally introduce \gim, before detailing its mechanisms. 
Given a retweet cascade $\His_\zeta(T)=\{v_1, v_2, \ldots\}$, \gim quantifies the social influence of a tweet $v_i$ as:
\begin{equation} \label{gim:eq:gim-expectation}
    \varphi_\gamma(v_i) = \sum_{ \mathcal{G} \in \Upsilon}  \sum_{t_j > t_i}
        \underbrace{\mathbb{P}_\gamma(\mathcal{G}_{i \rightarrow j})}_{\text{Conductance}} \quad
        \underbrace{\Psi(\mathcal{G}_{i \rightarrow j})}_{\text{Capital Distrib.}}
        \enspace,
\end{equation}
where $\Psi(\mathcal{G}_{i \rightarrow j})$ is the \socialcapital allocated to $v_i$ along the endorsement pathway from $v_j$, and $\mathbb{P}_\gamma(\mathcal{G}_{i \rightarrow j})$ is the conductance-mediated probability that a path exists between $v_i$ and $v_j$ in $\mathcal{G}$ (defined in \cref{gim:subsec:conductance}).
$\Psi$ is defined at the level of edges, preserving the efficiency of the \hawkesinf computation (see \cref{gim:subsubsec:efficient-computation}).
The influence of a user is the average influence of their tweets.

\subsection{Conductance}
\label{gim:subsec:conductance}

\revEPJA{Conductance models how readily influence flows between two users.
Just as physical materials conduct electricity or heat with varying efficiency, social relationships conduct influence to varying degrees: a user is more likely to be influenced by someone they are similar to or connected with than by a stranger.
The conductance of an edge modulates the probability of influence transmitted along that edge (see \cref{gim:eq:pij}).}

Quantitative models often assume that social ties are the primary influence channels, while other conductive channels (like homophily) are underexplored.
Cognitive science \cite{lin2018intergroup} and social psychology \cite{cialdini2001science, newcomb1953approach, centola2011experimental} literatures suggest that some relationships are more influential than others.
Notably, people in the same community (or who share  similarities) are more influential to each other.
Conductance assumes that different types of relations between users propagate influence more effectively (e.g., one might be more influenced by close family than by distant work colleagues);
this makes some people more likely sources of influence than others. 
Intuitively, the social system propagates influence similarly to physical materials conducting electricity or heat.
The conductance of a social connection modulates the likelihood of adopting an opinion.
The higher the conductance, the more likely the receiver will adopt the opinion of the emitter.
The influence conductance encapsulates user relationships;
by measuring which links are more conducive to influence, we can infer more realistic endorsement pathways.

The conductance mechanism modulates the likelihood of cascade pathways using user lexical and following similarity, and relationship ties.
We denote as $\gamma_{i,j}$ the conductance of an edge $(v_i, v_j)$, and define the updated probability that $v_j$ is a direct offspring of $v_i$ as
\begin{equation}
    \label{gim:eq:pij}
    p'_{ij} = \frac{\zeta_i^b \phi(t_j - t_i)\gamma_{i,j}}{\mu(t) + \sum_{t_k < t_j} \zeta_i^b \phi(t_j - t_k)\gamma_{k,j}}, t_i < t_j.
\end{equation}

We consider two choices for conductance: \emph{topological} (users' social network) and \emph{homophilic} (users' similarity with others).
We further operationalize homophilic conductance, using two lenses: \emph{following} and \emph{lexical}.

\sh{Topological conductance} assumes influence flows between users who are connected in the social graph (i.e., the follower relationship).
Each valid edge $(v_i, v_j), \text{ with } t_i < t_j$, has a baseline conductance $\beta_{top}$, regardless of whether $u_i$ is connected to $u_j$, accounting for alternative influence conduits (such as news feeds or users following topics and hashtags).
Formally, we define the topological conductance $\gamma^{top}_{ij} \in [0, 1]$ as $\gamma^{top}_{ij} := \beta_{top} + (1-\beta_{top}) a_{ij}$,
where $a_{ij} = 1$ if $u_j$ follows $u_i$ and 0 otherwise.

\sh{Homophilic conductance} of an edge, models the connection between similarity and influence, i.e., people similar to us influence us more~\cite{crandall2008feedback,goel2016social}.
For each user $u_i$ we first build a user representation $h_i \in \mathbb{R}^{n}$.
Next, we quantify $\gamma^{hom}_{i,j}$ the homophilic conductance between two users using the cosine similarity between their user representations plus the baseline conductance $\beta_{hom}$.
Formally, $\gamma^{hom}_{i,j} := \beta_{hom}  + (1 - \beta_{hom}) cosine(h_i,h_j)$.

The \emph{following lens} leverages the observation that similar people consume similar content.
On social media, following popular users is akin to consuming content.
Accordingly, we measure the similarity between two users based on whether they follow the same people.
For the following lens, we identify the $1000$ most followed users, and collect the followees of all the users in our dataset (i.e., users they follow).
We represent a user $u_i$ as $h_i \in \mathbb{R}^{1000}$, where $h_i[j] = 1$ if $u_i$ follows the $j$th most followed user ($h_i[j] = 0$ otherwise).

The \emph{lexical lens} exploits the insight that similar people use similar language.
The vocabulary and language style of users can be a strong indicator of their community, and we measure the similarity of users based on their choice of language.
For the lexical lens, we construct user documents by concatenating the user tweets;
we represent them using TF-IDF (Term Frequency-Inverse Document Frequency);
and a feature hashing dimensionality reduction technique~\cite{moody1989fast}.
Finally, we represent each user $u_i$ as $h_i \in \mathbb{R}^{1,048,576}$.
Note that the homophilic conductance with the lexical lens does not require knowledge of the user following graph, which can be prohibitively expensive to obtain for Twitter.  

Related to our work, \citet{centola2011experimental} characterize the relationship between homophilic ties and the influence phenomena, however their network is artificial and they illustrate only one homophilic lense.

\revEPJA{We note that other attributes---such as partisanship, gender, or race---may serve as stronger indicators of homophily than the lexical and following lenses used here.
We opt for lenses that are readily computable at scale without requiring sensitive demographic data, and leave the incorporation of richer sociological features into the conductance mechanism to future work.}

\subsection{\Socialcapital Distribution}
\label{gim:subsec:inf-cap-dist}
The endorsement pattern is not a sufficient explanation of the ability of individuals to exert influence.
Sociologists have long pointed to the importance of weak-ties \citep{granovetter1983strength} and the accumulation of social capital by \emph{bonders} \citep{dekker2003social}.
Consider the example in \cref{gim:fig:teaser}.
Alice influences Bobbie, who influences several people;
should Alice (the initiator) or Bobbie (the connector) be considered more influential?

Psychologists have recognized that particular characteristics in individuals are correlated with influence, namely authority \cite{milgram1978obedience}, attractiveness \cite{horai1974effects}, likeability and others \cite{cialdini2001science}.
We define \emph{\socialcapital} as the congruence of characteristics that enable the exertion of influence.
We propose that users possess latent \socialcapital, such that an emitter with higher \socialcapital exerts more influence. 
We posit that a distribution mechanism that explains endorsement pathways, must measure this \socialcapital.
\revEPJA{Concretely, we model \socialcapital as a quantity that accumulates through the endorsements a user receives: when a user is retweeted, a portion of the retweeter's capital flows upstream to them.
Users who trigger large or far-reaching cascades accumulate more capital, and thus are estimated to be more influential.}
The question remains, how do we allocate \socialcapital to explain the exerted influence along an inferred endorsement pathway?

We propose a \emph{\socialcapital distribution} that leads to the accumulation of \socialcapital and offers a post-hoc explanation for the inferred endorsement patterns. 
The mechanism transfers a proportion of a node's (tweet's) \socialcapital to its parent in the information cascade, allowing upstream and highly connected nodes to accumulate \socialcapital (which translates to influence).
The distribution mechanism that we propose here aims to explain the endorsement pathways inferred from the data,
and it is related to the concept of value-allocation schemes \cite{shapley1997value, ghaffar2020structural}.
Several studies \cite{jackson2003strategic,ghaffar2020structural,subbian2014finding} have utilized allocation schemes to recognize the role of \emph{bonders} by assuming benefits, generated by a node, decay with distance within a social graph.
We migrate this intuition to a directed diffusion scenario.
Intuitively, the allocation rewards users responsible for bonding (i.e. connectors), and those who reach distant communities (i.e. initiators).

We construct the \socialcapital distribution as follows.
Whenever the user $u_i$ directly influences $u_j$ (i.e., $v_j$ is a direct offspring of $v_i$ in an endorsement pathway), $u_j$ transfers a portion of their \socialcapital to $u_i$ (denoted as $\pi_{ij}$).
Each tweet is endowed with $1$ \socialcapital for participation; it pays a proportion $\alpha \in (0,1)$ of all its capital to its parent (if they exist) and keeps $(1-\alpha)$. 
Formally:
\begin{equation*}
    \pi_{ij} = \left\{ \begin{array}{lll}
        \alpha, & 1 \leq i < j, & \textcolor{gray}{j \text{ transfers } \alpha \text{\% capital to } i,}\\
        (1-\alpha), & i = j \neq 1, & \textcolor{gray}{j \text{ keeps } (1-\alpha) \text{\% capital},}\\
        1, & i=j=1,& \textcolor{gray}{\text{The initiator does not transfer.}}
    \end{array} \right.
\end{equation*}
\cref{gim:fig:teaser} illustrates the mechanism.
Charlie passes Bobbie $\alpha$ of her endowed capital (keeping $1-\alpha$), and Bobbie receives capital from her endowment ($1$), her three direct influencees ($3\alpha$), and one indirect influencee ($\alpha^2$).
She keeps $(1-\alpha)$ of this sum.
The capital distribution of a path $\mathcal{G}_{i \rightarrow j}$ is
$\label{gim:eq:capital-distrib-path} \Psi(\mathcal{G}_{i \rightarrow j}) = \prod_{(k,l) \in E(\mathcal{G}_{i \rightarrow j})} \pi_{kl},$
and a user's social influence is proportional to the total \socialcapital they accumulate via the capital distribution mechanism. The scheme naturally conserves total value, which is equal to the number of participants.

\subsection{Iterative Computation}
\label{gim:subsubsec:efficient-computation}
\gim can be computed efficiently by extending the pairwise influence $m_{ik}$ (introduced in \cref{gim:sec:preliminary}) to incorporate the concepts of conductance and \socialcapital distribution.
Formally,
    $
        \label{gim:eq:gim-mik}
        m_{ik} = \sum_{j=i}^{k-1} m_{ij}p'_{jk} \pi_{jk} , i<k \enspace,
    $
where $m_{ik}=p'_{ik}\pi_{ik}$ when $i=k$, and $m_{ik} = 0$  when $i>k$.
Consequently (and similar to the \hawkesinf), we obtain $\varphi_\gamma(v_i) = \sum_{k=i}^n m_{ik}$.
The full derivation of the latter, from \cref{gim:eq:gim-expectation} via \cref{gim:eq:pij,gim:eq:gim-mik}, is shown in the online appendix~\citep{appendix}.
\gim recursively generates all possible paths in the same way as the Hawkes-modeled Influence~\citep{rizoiu2018debatenight}, which allows an efficient iterative algorithm of temporal complexity $O(n^3)$.
Visibly, the \hawkesinf~\citep{rizoiu2018debatenight} is a special case of \gim, with $ \pi_{jk} = \gamma_{jk} = 1, \forall j, k$.

\section{\gim Evaluation}
\label{gim:subsec:gim-evaluation}
Having developed the theoretical foundation for our Generalized Influence Model, we now turn to its empirical evaluation.
This section addresses three key questions: (1) what combination of conductance and distribution mechanisms yields optimal performance? (2) how does \gim compare to existing influence estimation methods? and, (3) can \gim overcome the known biases of traditional influence metrics?

\sh{Ground Truth Influence.}
Social influence is inherently difficult to quantify, particularly in online settings where direct behavioral change is challenging to observe.
\revEPJA{By \emph{ground truth influence} we mean empirically measured social influence recovered from crowdsourced pairwise comparisons of Twitter/X profiles.}
We utilize our previously developed empirical influence ground truth \mbox{\cite{empiricalMeasurement}} which employs a human-in-the-loop active learning approach to measure peer-perceived influence.
The method leverages crowdsourcing workers who perform pairwise comparisons between users, determining which is more influential.
\revEPJA{For each comparison, workers are shown the two users' profiles, recent tweets, and links to their public pages, and are asked to judge which user is more influential.
The limitations of this approach are explored in detail in \citet{empiricalMeasurement}.}
To optimize the required number of comparisons, the method implements a Quicksort-based active learning algorithm that selects the most informative pairs to compare, reducing the quadratic complexity of naive approaches to a loglinear one.
The comparisons are then used to fit a Bradley-Terry model, which produces a complete ranking of social influence scores.
\revEPJA{By constructing rankings from pairwise comparisons, this approach avoids the ordinality distortions that arise when psychometric constructs---which lack standardized units of measurement---are coded as numerical scores \citep{carpentras2023psychometric}.}
The resulting dataset consists of influence rankings for $500$ Twitter users selected from the \#ArsonEmergency dataset ~\cite{graham2020bushfires}--- containing Twitter discussions about the Australian bushfires collected between November 2019 and January 2020.
This empirical influence measure has been validated by showing strong correlation with the Big Two of social cognition (agency and communion), theoretical determinants of social influence.
We utilize this empirical influence ranking to calibrate and evaluate \gim.

\sh{Evaluation metrics.}\label{gim:para:metrics}
We evaluate \gim ranking against the ground truth (described above) using two measures: 
\ndcgauc (see next) and MAPE.
The information retrieval literature uses the \emph{Normalised Discounted Cumulative Gain} (NDCG) to measure the overlap between two rankings.
It privileges the correct ranking of the top-ranked positions and discounts errors in the lower rankings.
Applied to influence, NDCG@k aims to order the $k$ most influential users correctly.
We compute the Area Under Curve for NDCG@k (\ndcgauc) by varying $k$ and producing a single metric value.
We also compute the Mean Absolute Percentage Error (MAPE) of the difference in the ranking percentiles for each target.
Having established our evaluation methodology, we next introduce the baseline methods against which we compare \gim.

\sh{Influence ranking baselines.}
We compare \gim to four baselines.
Two baselines are widely used heuristics; \emph{PageRank}~\cite{page1999pagerank,xiang2013pagerank} assumes influence flows via random-walks on constructed social graphs (here the follower network) and \emph{retweet influence}~\cite{cha2010measuring}, counts the retweets of a user's authored tweets. 
They are centrality- and feature-based approaches, respectively.
The other two baselines are purpose-built state-of-the-art influence estimators: \emph{\hawkesinf}~\citep{rizoiu2018debatenight} and \emph{ProfileRank}~\cite{silva2013profilerank} (a PageRank variant).
Next, we conduct a comprehensive search to identify the optimal configuration of our proposed model.

\begin{figure*}[!tbp]
    \centering
    \includegraphics[width = \textwidth]{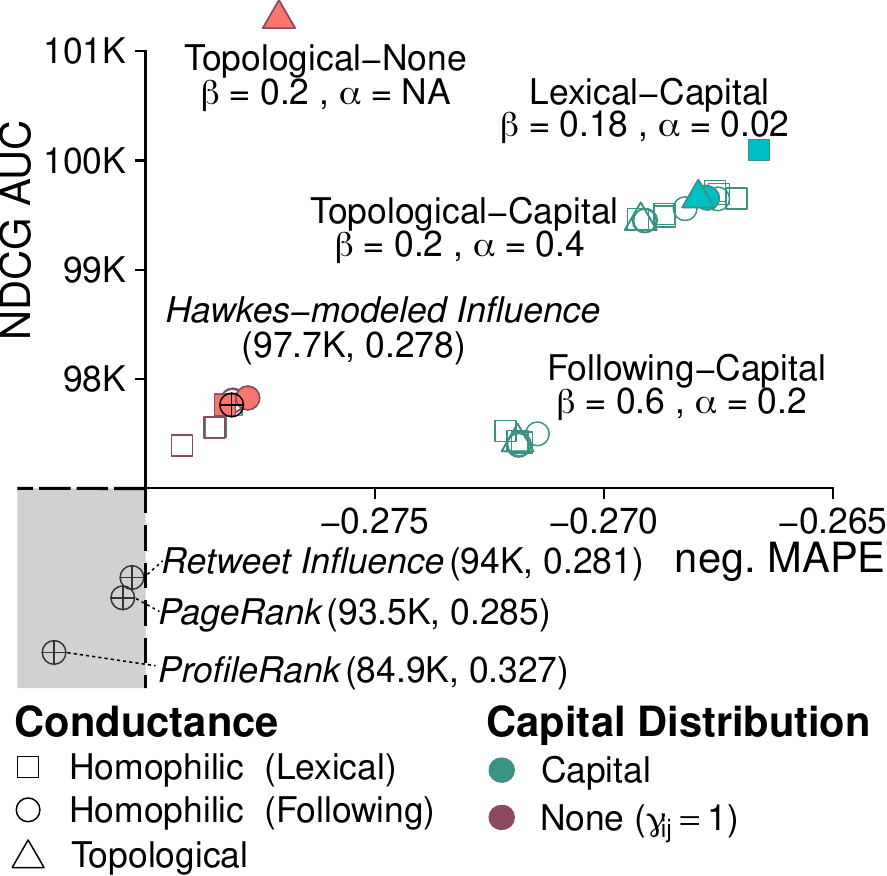}\caption{        
        \emph{Evaluate \gim against the ground-truth in the space of \ndcgauc (y-axis) and negative MAPE (x-axis).}
        Higher \ndcgauc and more negative MAPE indicates a better performing model.
        The solid shapes are the best models for each combination (conductance--capital distribution).
        The empty shapes show the Pareto-dominated models in each combination, obtained via grid search in the space $(\beta, \alpha)$.
        The circle-crosses denote the baselines: \hawkesinf baseline~\citep{rizoiu2018debatenight}, PageRank~\cite{page1999pagerank}, Retweet Influence~\cite{cha2010measuring}, and ProfileRank~\cite{silva2013profilerank}. 
        Note, the \textcolor{gray}{gray box} is not to scale, and the coordinates for baselines are shown in brackets.
    }
    \label{gim:subfig:orthogonal-influence-plot}
\end{figure*}

\sh{\gim search.}
For each combination of conductance (topological, lexical, and following) and distribution mechanism (\socialcapital, and none), we perform a grid search over the hyper-parameters $\beta$ (conductance) and $\alpha$ (distribution).
At each grid point, we compare the influence scores obtained by \gim against the ground truth empirical influence ranking. \cref{gim:subfig:orthogonal-influence-plot} shows the baselines and \gim with several conductance-distribution combinations in the space of the performance measures: negative MAPE (x-axis) and \ndcgauc (y-axis) (the top-right corner optimizes both measures).

Analysis of the experimental results reveals three key observations.
First, \gim consistently Pareto-dominates (i.e., outperforms) all baselines for almost every hyperparameter combination, showing that our psychosocial-inspired mechanisms render automatic influence quantification closer to the human judgment.
Among baselines, the next best performing is the \hawkesinf, followed by PageRank, retweet influence, and the purpose-built ProfileRank.
Second, we observe that the homophilic conductances (i.e., lexical and following) typically outperforms the topological conductance, and between homophilic conductances lexical outperforms the following conductance.
Third, only two models are not Pareto-dominated: the topological-none (best \ndcgauc) and the lexical-\socialcapital (best neg. MAPE).
Notice, however, that the topological conductance requires recovering the follower network.
In practice, this is prohibitive for large datasets (such as the \covid dataset) due to rate limitations of the Twitter/X API.
Based on these findings, in our analysis in \cref{gim:subsec:occupational-application} we use the homophilic lexical conductance ($\beta = 0.18$) with the \socialcapital distribution ($\alpha = 0.02$).
The $18\%$ baseline conductance indicates that the accounted channels do not explain a relatively large proportion of conductance.
Note that while passing $2\%$ of the \socialcapital to the parent might not seem much, this adds up for nodes with high degrees (particularly given the longtail distribution of follower count~\cite{Wu2019}).
Having established \gim's superior performance against existing models, we now examine its ability to address a key limitation of traditional influence metrics.

\begin{figure*}[tbp]
    \newcommand\myheight{0.28}
    \centering
    \includegraphics[height=\myheight\textheight]{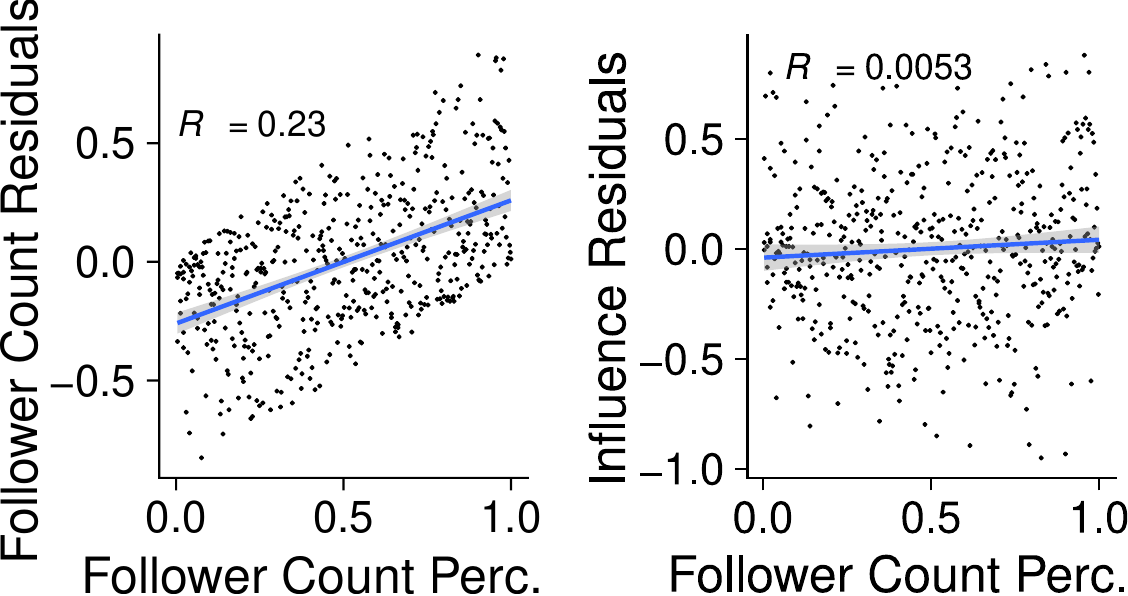}
\caption{
        Residuals (relative to the empirical follower ranking) for the follower count (left) and \gim (right) against the follower count (x-axis).
    }
    \label{gim:subfig:debiasing-plots}
\end{figure*}

\sh{Debiasing the follower count.}
The follower count is widely used as a proxy for influence~\cite{cha2010measuring,frantz2009robustness, riddell2017most}; 
however, it has been repeatedly shown to be biased~\cite{bakshy2011everyone,smith2018influence, romero2011influence}. 
We examine whether \gim can overcome this known limitation.
\cref{gim:subfig:debiasing-plots}(left) shows that the follower count residuals---the difference between the follower count percentile and the empirical score percentile---are positively correlated with the follower count percentile ($R^2 = 0.48$).
In other words, the follower count overestimates the influence of the highly followed users and underestimates the lowly followed users.
In contrast, \cref{gim:subfig:debiasing-plots}(right) shows that \gim residuals are not correlated with the follower count percentile ($R^2 = 0.073$).
That is, \gim is an unbiased influence estimator with respect to the follower count.
Having demonstrated \gim's effectiveness at quantifying influence in an unbiased manner, we next proceed to apply it to real-world data to investigate the relationship between social class, expertise, and influence in online discourse.

\section{Social Class, Expertise, and Influence during COVID}
\label{gim:subsec:occupational-application}

Having established GIM as an effective influence quantification model that outperforms state-of-the-art approaches and corrects inherent biases, we now apply this tool to examine fundamental sociological hypotheses.
Sociologists have long espoused that power and influence are primarily related to social class and expertise \mbox{\cite{kraus2019social}}. 
The influence of experts, and trust in their advice, is important for the optimal functioning of society. 
Failure to adhere to advice can lead to unnecessary death (e.g., vaccine hesitancy) and existential threats (e.g., climate change). 
Social class is often nebulously defined but is related to wealth, occupation, and subculture.
In our increasingly egalitarian societies, occupation provides a suitable proxy.
\revEPJA{We choose occupation as our primary lens not because it is the strongest correlate of homophily---attributes such as partisanship, gender, or race may exhibit stronger homophilic ties---but because of its theorised relationship to influence through authority, prestige, and expertise \citep{cialdini2001science, kraus2019social}.
Moreover, occupation is among the most accessible and verifiable attributes at scale, as users frequently self-report it in their social media profiles.}
The COVID-19 pandemic presents an ideal context for investigating these relationships, as it constitutes a crisis where expert communication should be paramount for public safety.
Ideally, influence, expertise, and social class are aligned; however, this does not seem to have been the case as \citet{Bailo2023} observed that far-right accounts, which had been peripheral during previous crisis events, managed to assume more central positions in online COVID-19 discussions.

In this section, we leverage our validated \gim framework to empirically examine the interplay between social influence, expertise, and occupational class during the COVID-19 pandemic.
We first introduce the COVID dataset and use \gim to compute the influence of all users in the \covid dataset. 
Next, we develop methods to quantify two key variables: users' occupations as proxies for social class, and the veracity of information they spread as an indicator of expertise.
We extract users' occupations and the veracity of the information they spread, and we tabulate their influence and veracity against their occupation.
Finally, we analyze the relationships between these variables to assess whether influence patterns align with theoretical expectations about expertise and social class.

\sh{Dataset.}
The dataset was collected from Twitter/X in the context of the COVID-19 pandemic. 
The \covid dataset was constructed using the keyword \textit{covid19} during August 2020 and contains $143,356,591$ tweets by $21,527,913$ users.

Note, this dataset is different from the \#ArsonEmergency dataset, we used to tune \gim.

\sh{Determine the occupations of Twitter/X users.}
To establish occupational classifications as proxies for social class, we implement a systematic approach for identifying users' professions \cite{Kern2019}.
We match user occupation against the Minor Group Occupational Classes of the O*NET occupational taxonomy \citep{onet} using textual fuzzy matching \citep{turrell2018using}.
O*NET described users better than other taxonomies investigated.
We search each user's Twitter/X description and select the first matched occupation (following \citet{sloan2015tweets})---assuming people list their actual occupation first, before hobbies and other information.
We validated the classifier on $100$ labeled users, where ground-truth labels were derived by two annotators, with disagreement resolved by discussion. 
The classifier has a mean macro-F1 of $0.54$, comparable to the classification performance reported by literature \cite{mukherjee2021determining}.

\sh{Quantifying Expertise.}
We assume that expertise is correlated with the veracity (i.e., quality) of the content users share.
To operationalize expertise in the online domain, we develop a method for measuring the quality of information shared by users.
We use the links and tweets that users share to quantify the veracity of the information they spread. 
We follow prior research~\cite{singh2020understanding} and compute a veracity score for the domain names of the URLs.
First, we extract from the CoAID dataset \cite{cui2020coaid} all the links and tweets associated with \emph{true} and \emph{fake} information related to COVID-19.
The dataset contains full URLs, and some URL domains appear many times.
For each URL domain within CoAID, we count the true ($\#R$) and fake ($\#F$) entries recorded in the dataset.
Finally, we generate for each domain a normalized score as $\frac{\#R - \#F}{\#R + \#F}$.
The score is between $-1$ (domain is fully unreliable and spreads misinformation) and $1$ (fully reliable).
Furthermore and similar to \cite{singh2020understanding}, we curate a set of \emph{high-quality health sources} (HQHS) from prominent health websites (e.g., CDC, WHO, and Mayo Clinic), and medical journal websites (e.g., The Lancet, and Nature).  
We compute the veracity score of a post with a link as follows:
(1) $-1$ or $1$ if the full link appears in CoAID as fake or true, respectively;
(2) $1$ if the link domain appears in the HQHS set;
(3) the domain's veracity $\in [-1,1]$, if neither (1) nor (2) applies.
A user's veracity score is the mean veracity of the posts they share.

\begin{figure*}[tbp]
    \newcommand\myheight{0.23}
    \centering
    \subfloat[]{
        \includegraphics[height=\myheight\textheight]{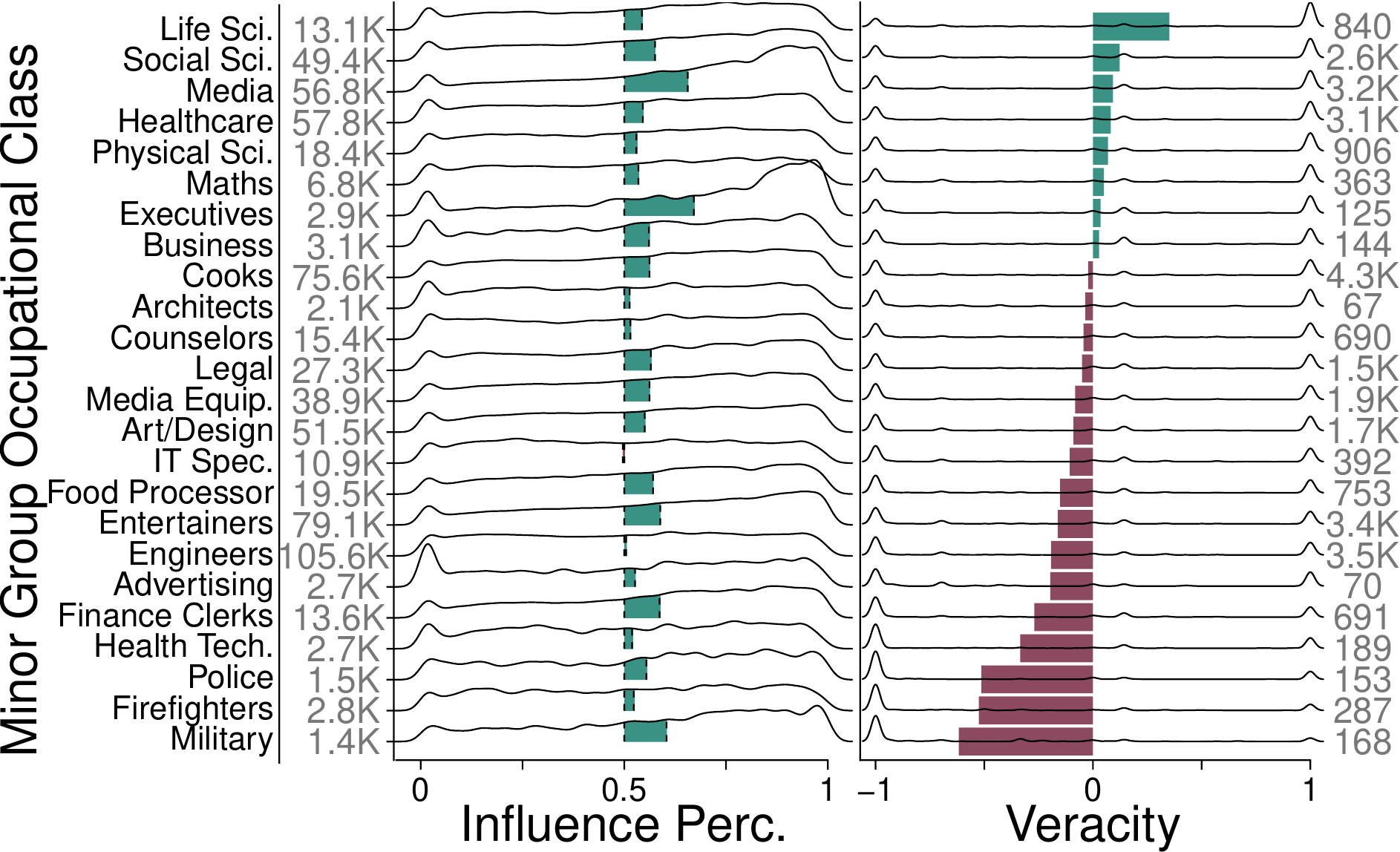}\vspace{-2mm}
        \label{gim:subfig:occupation_distributions}
    }\subfloat[]{
        \includegraphics[height=\myheight\textheight]{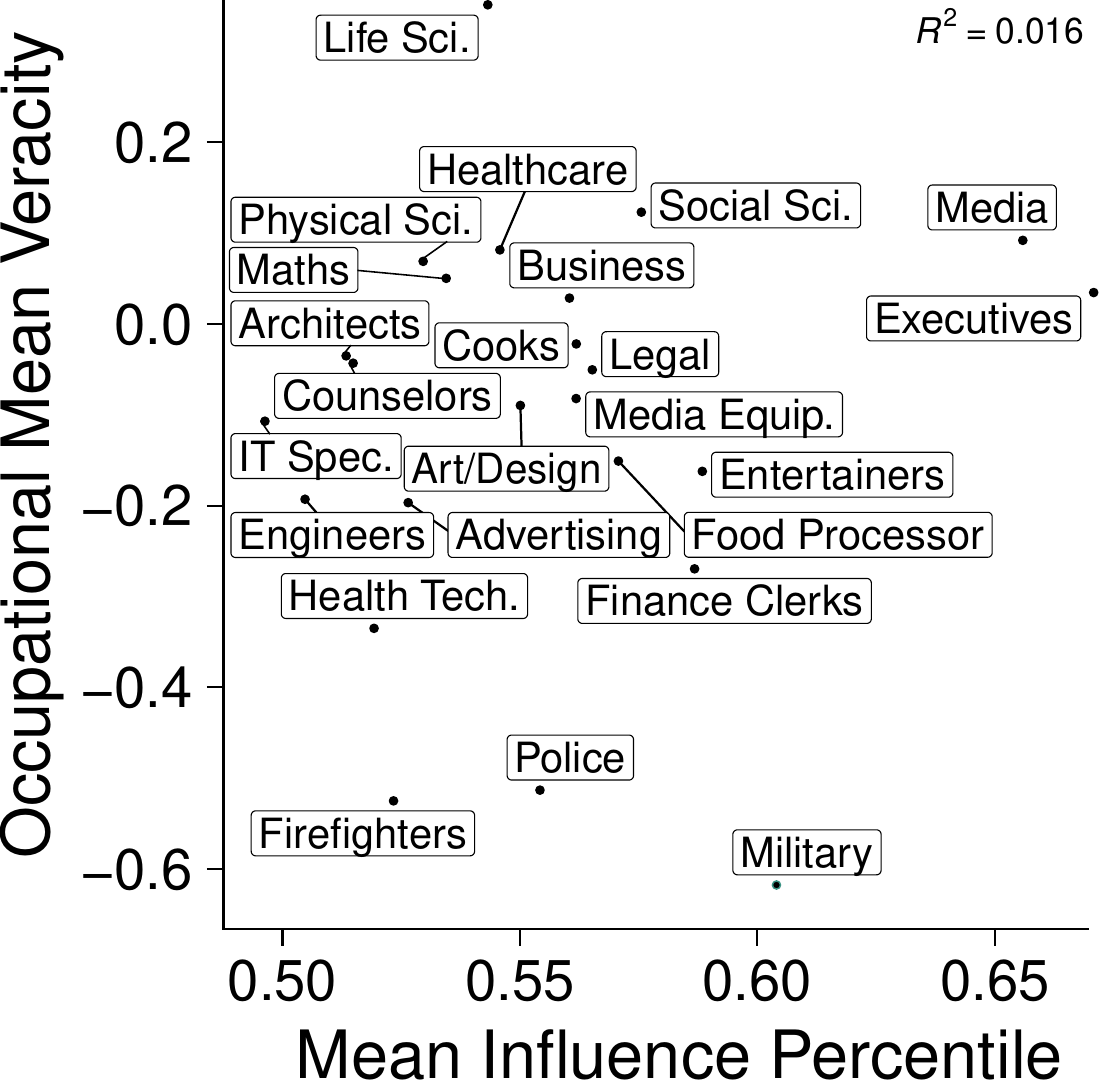}\vspace{-2mm}
        \label{gim:subfig:occupation_scatter}
    }\vspace{-3mm}
    \caption{
        (a)(left panel) The influence distribution for the O*NET Minor Group occupations with more than 1,000 users in \covid (number of users shown on the left).
        (a)(right panel) The veracity distribution of the same occupations (number of spreaders shown on the right).
        Color bars show the difference between the occupation mean and the distribution's center ($0.5$ for influence and $0$ for veracity).
        (b) The mean veracity (y-axis) and mean influence percentile (x-axis) of occupations.
    }
    \label{gim:fig:set-two}
    \vspace{-3mm}
\end{figure*}

\sh{The influence and veracity of occupations.}
Having quantified influence, occupation, and information veracity, we now analyze how these variables interact across different professional groups during the pandemic.
\cref{gim:subfig:occupation_distributions} shows the distribution of user influence and user veracity scores, for occupations with more than a thousand users in the \covid dataset.
\replaceEPJA{Our model estimates that \textit{Executives}, the \textit{Media}, \textit{Entertainers} and the \textit{Military} are among the most influential occupations.}{\textit{Executives}, the \textit{Media}, \textit{Entertainers} and the \textit{Military} yield among the highest online influences.}
This result is hardly surprising for the former three.
Notably, \textit{Media} and \textit{Entertainers} are not only influential but have prominent online presences, perhaps on account of their attention-related business models.
The latter (\textit{Military}) is an occupation with high bipartisan support and respect in the US -- home of most English-speaking Twitter/X users. 
Examining the veracity distributions reveals important patterns in information quality across occupational groups.
For all occupations, \cref{gim:subfig:occupation_distributions} shows that the veracity score has a bimodal distribution.
One mode is around $-1$ (the users who spread mainly misinformation) and another at $1$ made of users who spread high-quality information.
Users typically do not engage with both types of information, probably due to homophily and online polarization.
As all occupational subpopulations contain both types of users, the mean occupation veracity (colored horizontal bars in the right panel of \cref{gim:subfig:occupation_distributions}) represents the ratio between misinformation and high-quality information spreaders within an occupation.
\replaceEPJA{Contrary to theoretical expectations about expertise and information quality, our analysis suggests that several occupations traditionally associated with authority spread substantial misinformation.}{Contrary to theoretical expectations about expertise and information quality, we find that several occupations traditionally associated with authority spread substantial misinformation.}
Surprisingly, we observe that a significant proportion of \textit{Military}, \textit{Firefighters} and \textit{Police} users spread misinformation. 
Closer investigation shows they frequently share from controversial publishers, such as \textit{foxnews.com}, \textit{zerohedge.com}, and \textit{breitbart.com}. 
To better understand the relationship between influence and information quality, we examine their correlation across occupational groups.
\cref{gim:subfig:occupation_scatter} shows the scatterplot of the occupations, in the space of mean influence (x-axis) and mean veracity (y-axis).
First, we observe that the two quantities are uncorrelated ($R^2 = 0.016$).
\replaceEPJA{This suggests a tension with theoretical expectations that expertise (measured as information quality) would align with influence.}{This finding contradicts theoretical expectations that expertise (measured as information quality) would align with influence.}
We also see that \textit{Life Scientists} (including epidemiologists) and \textit{Social Scientists} typically spread high-quality information; however, our model estimates limited influence for these groups.
\textit{Healthcare} workers are moderately influential and have a good proportion of high-quality information spreaders; 
however, \textit{Healthcare Technicians} including paramedics and medical technologists, have a significant proportion of misinformation spreaders.
\revEPJA{We note that these findings reflect users active in the tracked COVID-19 discussions on Twitter/X, and that keyword-based data collection may bias the observed user sample.}

\section{Discussion}
\label{gim:sec:Conclusion}

Existing methods for online influence measurement rely on heuristic definitions and fail to model complex phenomena.
In this work, we have developed an online social influence framework.
Our conductance mechanism departs from traditional approaches in the literature which frequently considers the social network (i.e. the topological lens) as the primary channel of influence \cite{kempe2003maximizing,du2013scalable}. 
As online interactions are increasingly mediated by recommender systems, web search, and cross-platform engagement, modeling the similarity between users (i.e. the homophilic lens) provides a simple and effective method for modeling influence channels.
Our findings demonstrate that lexical features are readily attainable and perform influence modeling well (see \cref{gim:subfig:orthogonal-influence-plot}).
The conductance mechanism we have introduced extends beyond the current application and could be used to mediate diffusion models or as a more general conception of the social graph, for social tasks (i.e. recommendation) or social network analysis.

\revEPJA{A foundational challenge in network diffusion research is disentangling peer influence from homophily: observed behavioural contagion may reflect genuine peer-to-peer influence or may arise because similar individuals independently adopt similar behaviours \citep{aral2009distinguishing}.
Prior work, such as \citet{aral2009distinguishing}, treats influence and homophily as independent, competing causes and decomposes the observed effect additively.
\gim takes a fundamentally different position: rather than treating homophily as a rival explanation to be separated from influence, the conductance mechanism models homophily as the \emph{channel through which} influence propagates.
Higher similarity between users increases conductance, making influence transmission more likely along homophilic ties.
We acknowledge that fully disentangling causal peer influence from homophily-driven correlation requires experimental or quasi-experimental designs beyond the scope of this observational study.}

\revEPJA{A growing body of evidence suggests that the spread of controversial and political content on social media follows complex contagion dynamics, where adoption requires reinforcement from multiple exposures rather than a single contact \citep{Romero2011, Bakshy2009, Monsted2017, Guilbeault2018}.
While \gim employs the independent cascades assumption \citep{kempe2003maximizing} for computational tractability, the underlying Hawkes process captures a form of social reinforcement consistent with complex contagion: the conditional intensity at any time is the sum of contributions from \emph{all} prior events in the cascade, so the likelihood of a new event reflects cumulative exposure rather than a single contact (see \cref{gim:app:gim-derivation}).
However, \gim does not model the non-linear threshold effects that are also associated with complex contagion, such as wide bridges and complex paths \citep{Guilbeault2021}.
Incorporating such non-linear effects would significantly impact the scalability of the model, and we leave this extension to future work.}

The \socialcapital distribution mechanism in our \gim framework provides an intuitive approach for allocating \socialcapital to explain influence (as perceived by peers). 
This approach significantly improves upon the arbitrary allocations of prior influence models and integrates naturally with the scalable procedure for estimating influence over the stochastic diffusion graph.

The literature constructed the theoretical relationship between social influence and social class, with occupation serving as a key indicator of social class via educational attainment, income, and prestige.
\replaceEPJA{Through our analysis, we have estimated occupational influence in discussions around the COVID-19 pandemic (see \cref{gim:subfig:occupation_scatter}).}{Through our analysis, we have measured occupational influence in discussions around the COVID-19 pandemic (see \mbox{\cref{gim:subfig:occupation_scatter}}).}
Our results reveal that certain aspects of social class can explain our results; for example, the influence of \textit{Executives} might be explained by their income and occupational prestige.
\replaceEPJA{Interestingly, our model suggests that individuals with high educational attainment are not influential in this context.}{Interestingly, we find that individuals with high educational attainment are not influential in this context.} 
Additionally, our findings suggest a significant determinant of influence, with both \textit{Media} and \textit{Entertainers}, having high influence. 
More broadly, the modeling of social influence can have broader implications for understanding how extreme opinions infiltrate mainstream discussions \citep{Kong2022}.

\revEPJA{We acknowledge several limitations of our occupational analysis.
While sociological theory posits a strong link between social class and influence, influence may also be related to other sociographic attributes beyond occupation.
We chose occupation for its theorised relationship to influence determinants (authority, prestige, expertise) and its accessibility and verifiability at scale.
Further empirical studies examining influence through additional sociographic lenses are warranted and left to future work.
\gim is designed to quantify the perceived influence of individual users; the occupational analysis aggregates these individual estimates into cohort-level averages and is therefore a downstream application.
The core model is rigorously evaluated against a human-derived ground truth, compared to four baselines across multiple mechanism combinations via grid search, and shown to debias the follower count metric (see \cref{gim:subsec:gim-evaluation}), characterising its expected performance on new data.
The occupational conclusions additionally depend on the observed user sample, which is shaped by keyword-based data collection and the heavy-tailed distribution of influence.
Future work could further strengthen these findings by empirically measuring the influence of occupational groups and comparing them against the model estimates.}

In conclusion, we have presented an online influence model that could be used to; provide insights into downstream modeling such as opinion dynamics, understand the spread of misinformation, or identify influential individuals in political/social campaigns.
\replaceEPJA{Our model suggests that \textit{Media}, \textit{Executives}, \textit{Entertainers}, and the \textit{Military} are among the most influential occupations; however the latter contain a large proportion of misinformation spreaders.
In contrast, pandemic experts (i.e. \textit{Life Scientists}) are estimated to have limited influence.
These findings raise questions about the effectiveness of information dissemination by experts in situations of crisis.}{Our findings reveal that \textit{Media}, \textit{Executives}, \textit{Entertainers}, and the \textit{Military} are the most influential; however the latter contain a large proportion of misinformation spreaders.
In contrast, pandemic experts (i.e. \textit{Life Scientists}) have only limited influence.
These results highlight the critical need for the amplification of our experts' content.} 
\section*{Acronyms \& Abbreviations}

\begin{acronym}[AAAAAA]
    \acro{COVID-19}{Coronavirus disease}
    \acro{GIM}{Generalized Influence Model}
    \acro{O*NET}{Occupational Information Network}
    \acro{NDCG}{Normalised Discounted Cumulative Gain}
    \acro{AUC}{Area under the curve}
    \acro{MAPE}{Mean Absolute Percentage Error}
    \acro{API}{Application Programming Interface}
    \acro{URL}{Universal Resource Locator}
    \acro{US}{United States of America}
\end{acronym}

\backmatter

\bmhead{Declarations}~\newline

\sh{Availability of data and material.}
The datasets analyzed and code required to reproduce the results are available in the GitHub repository, \url{https://github.com/behavioral-ds/generalized-model-of-social-influence}. Auxiliary code and/or data is available from the corresponding author upon reasonable request.

\sh{Competing Interests.}
There are no conflicts of interest for either of the authors, and both have agreed to this submission. 

\sh{Funding.}
This research was supported by an Australian Government Research Training Program (RTP) Scholarship, 
the Advanced Strategic Capabilities Accelerator (ASCA), 
the Australian Department of Home Affairs, 
the Defence Science and Technology Group,
the Defence Innovation Network, 
the National Science Centre, Poland (Project No. 2021/41/B/HS6/02798).

\sh{Author Contributions.}
Both authors contributed substantially to the manuscript's conceptualization, design, data collection, analyses, interpretation, and drafting. 

\sh{Acknowledgements.}
The authors extend their gratitude to Dr. Manuel Cebrian who, during an informal discussion over coffee more than a decade prior to this publication, hypothesized the existence of a conductance property in social networks analogous to electrical conductance in physical systems.

\begin{appendices}

\section{Complete derivation of \gim}
\label{gim:app:gim-derivation}
This section shows the complete derivation from \cref{gim:eq:gim-expectation} to \cref{gim:eq:gim-mik}, both shown in the main text.

We define the influence of a tweet given the diffusion scenario $\mathcal{G}$, as:
\begin{equation}
    \varphi(v_i|\mathcal{G}) = \sum_{v_j \in  V(\mathcal{G})} \theta(i,j | \mathcal{G}) \label{gim:eq:infl-in-a-scenario}
\end{equation}
where $V(\mathcal{G})$ is the set of nodes in $\mathcal{G}$, and we denote as $\theta(i,j | \mathcal{G}) := \Psi(\mathcal{G}_{i \rightarrow j})$ the \socialcapital distribution along a path.
We start from the definition of \gim given a retweet cascade (main text \cref{gim:eq:gim-expectation}):
\begin{align} 
    \varphi_\gamma(v_i) &= \sum_{ \mathcal{G} \in \Upsilon}  \sum_{t_j > t_i}
    \underbrace{\mathbb{P}_\gamma(\mathcal{G}_{i \rightarrow j})}_{\text{Conductance}} \quad
    \underbrace{\Psi(\mathcal{G}_{i \rightarrow j})}_{\text{Capital Distrib.}}
    \nonumber \\
    &= \sum_{ \mathcal{G} \in \Upsilon} \mathbb{P}_\gamma(\mathcal{G})\sum_{v_j \in  V(\mathcal{G})} \theta(i,j | \mathcal{G}) = \sum_{ \mathcal{G} \in \Upsilon} \mathbb{P}_\gamma(\mathcal{G})  \varphi(v_i|\mathcal{G}) \label{gim:eq:brute-user-infl}
\end{align}

Notably, due to the factorial number of diffusion scenarios in $\Upsilon$, computing the influence for each graph is intractable.

\textbf{Incremental construction of diffusion scenarios.}
We leverage the \textit{independent cascades assumption} (see \cref{gim:sec:preliminary}) to construct an efficient influence computation that overcomes intractablility.
The key observation is that each tweet $v_k$ is added simultaneously at time $t_k$ to all diffusion scenarios constructed at time $t_{k-1}$. 
$v_k$ contributes only once to the tweet influence of every tweet found on the path to which $v_k$ is attached. 
The tweet influence is computed incrementally by updating $\varphi(v_i), i < k$ at each time $t_k$. 
We denote by $\varphi^k(v_i)$ the value of tweet influence of $v_i$ after adding node $v_k$. 
As a result, we only track how the tweet influence increases over time steps, and we do not construct all valid diffusion scenarios.

\gim assumes that a user's tweet is influenced by one of the precedent tweets, chosen stochastically from a discrete distribution over the valid edges.
Alternatively, we can interpret that all previous tweets influence the new tweet proportionally to the same discrete distribution (a view inline with recent findings about influence and complex contagion).
Let $\Upsilon_{1:k-1}$ be the set of all possible diffusion scenarios at time $t_{k-1}$, and $\mathcal{G}^- \in \Upsilon_{1:k-1}$ be one such diffusion scenario, with the set of nodes $V^- = \{ v_1,v_2,\cdots,v_{k-1} \}$.
When $v_k$ arrives, it can attach to any node in $V^-$, generating $k-1$ new diffusion scenarios $\mathcal{G}_j^+$, with $V_j^+ = V^- \cup v_k$ and $E_j^+ = E^- \cup (v_j, v_k)$. We can write the set of scenarios at time $t_k$ as:
\begin{equation} \label{gim:eq:diff-scenario-increase}
	\Upsilon_{1:k} = \left\lbrace \mathcal{G}_j^+ \middle| \forall j < k, \forall \mathcal{G}^- \in \Upsilon_{1:k-1} \right\rbrace
\end{equation}

We write the tweet influence of $v_i$ at time $k$ as:
\begin{align}
    \varphi^k(v_i) = \sum_{\mathcal{G}^+ \in \Upsilon_{1:k}} \mathds{P}_\gamma(\mathcal{G}^+) \varphi(v_i|\mathcal{G}^+) =^{cf.~\eqref{gim:eq:diff-scenario-increase}} \sum_{\mathcal{G}^- \in \Upsilon_{1:k-1}} \sum^{k-1}_{j=1}\mathds{P}_\gamma(\mathcal{G}_j^+) \varphi(v_i|\mathcal{G}_j^+)  \label{gim:eq:infl-step-k}
\end{align}

\textbf{Attach a new node $v_k$.}
We concentrate on the right-most factor in Eq.~\eqref{gim:eq:infl-step-k} -- the tweet influence in scenario $\mathcal{G}^+_j$. 
We observe that the terms in Eq.~\eqref{gim:eq:infl-in-a-scenario} divide into two:
the paths from $v_i$ to all other nodes except $v_k$ (i.e. the old nodes) and the path from $v_i$ to $v_k$.
We obtain: 
\begin{equation*}
    \varphi(v_i|\mathcal{G}_j^+) = \sum_{\substack{v_l\in{\mathcal{G}_j^+}\\l>i, l\neq k}} \theta(i,l|\mathcal{G}_j^+) + \theta(i,k|\mathcal{G}_j^+)  
\end{equation*}

Note that a path that does not involve $v_k$ has the same \socialcapital contribution in $\mathcal{G}_j^+$ and in its parent scenario $\mathcal{G}^-$, i.e. $\theta(i,l|\mathcal{G}_j^+) = \theta(i,l|\mathcal{G}^-), \text{ for } l > i, \text{and } l \neq k$.
We obtain
\begin{align}
    \varphi(v_i|\mathcal{G}_j^+) 		&= \sum_{\substack{v_l\in{\mathcal{G}^-}\\l>i}} \theta(i,l|\mathcal{G}^-) + \theta(i,k|\mathcal{G}_j^+) =
^{cf.~\eqref{gim:eq:infl-in-a-scenario}} \varphi(v_i | \mathcal{G}^-) + \theta(i,k|\mathcal{G}_j^+) \label{gim:eq:infl-step-k+1}
\end{align}
Combining Eq.~\eqref{gim:eq:infl-step-k} and~\eqref{gim:eq:infl-step-k+1}, we obtain:
\begin{align}
&\varphi^k(v_i) = \sum_{\mathcal{G}^- \in \Upsilon_{1:k-1}}\sum^{k-1}_{j=1} \mathds{P}_\gamma(\mathcal{G}_j^+) \bigg[ \varphi(v_i | \mathcal{G}^-) + \theta(i,k|\mathcal{G}_j^+) \bigg] \nonumber \\
&= \underbrace{\sum_{\mathcal{G}^- \in \Upsilon_{1:k-1}} \varphi(v_i | \mathcal{G}^-) \sum^{k-1}_{j=1}\mathds{P}_\gamma(\mathcal{G}_j^+) }_{A}  
+ \underbrace{\sum_{\mathcal{G}^- \in \Upsilon_{1:k-1}}\sum^{k-1}_{j=1} \mathds{P}_\gamma(\mathcal{G}_j^+) \theta(i,k|\mathcal{G}_j^+)}_{m_{ik}} \label{gim:eq:two-parts}
\end{align}

\textbf{Tweet influence at previous time step $t_{k-1}$.} 
Given the definition of $\mathcal{G}_j^+$ in \cref{gim:eq:diff-scenario-increase} and the independant cascades assumption, we obtain that $\mathds{P}_\gamma(\mathcal{G}_j^+) = \mathds{P}_\gamma(\mathcal{G}^-)p'_{ij}$.
Consequently, part $A$ in \cref{gim:eq:two-parts} can be written as:
\begin{align}
	A 	&= \sum_{\mathcal{G}^- \in \Upsilon_{1:k-1}} \varphi(v_i | \mathcal{G}^-) \sum^{k-1}_{j=1} \mathds{P}_\gamma(\mathcal{G}^-)p'_{jk} \nonumber \\
		&= \sum_{\mathcal{G}^- \in \Upsilon_{1:k-1}} \varphi(v_i | \mathcal{G}^-) \mathds{P}_\gamma(\mathcal{G}^-) \sum^{k-1}_{j=1} p'_{jk} \stackrel{{cf.~\eqref{gim:eq:brute-user-infl}}}{=} \varphi^{k-1}(v_i) \label{gim:eq:past-infl}
\end{align}
$A$ is the tweet influence of $v_i$ at the previous time step $t_{k-1}$.
Note that $\sum^{k-1}_{j=1} p'_{jk} = 1$ because $v_k$ is necessarily the direct retweet of a previous nodes $v_j, j < k$ of the retweet cascade.

\textbf{Contribution of $v_k$.}
With $A$ being the influence of $v_i$ at the previous time step, intuitively $m_{ik}$ is the contribution of $v_k$ to the influence of $v_i$.
Knowing that:
\begin{align*}
	\mathds{P}_\gamma(\mathcal{G}_j^+) &= \mathds{P}_\gamma(\mathcal{G}^-)p'_{jk} \text{ and } \\
	\theta(i,k|\mathcal{G}_j^+) &= \Psi(\mathcal{\mathcal{G}}_{i \rightarrow j}) \pi_{jk} = \theta(i,j|\mathcal{G}^+_j) \pi_{jk} ^{cf. \ref{gim:eq:capital-distrib-path}} = \theta(i,j|\mathcal{G}^-) \pi_{jk}
\end{align*}
we write $m_{ik}$ as:
\begin{align*}
	m_{ik} &= \sum_{\mathcal{G}^- \in \Upsilon_{1:k-1}} \sum^{k-1}_{j=1} \mathds{P}_\gamma(\mathcal{G}^-)p'_{jk} \theta(i,j|\mathcal{G}^-) \pi_{jk} \nonumber \\
           &= \sum^{k-1}_{j=1} p'_{jk}\pi_{jk} \underbrace{ \sum_{\mathcal{G}^- \in \Upsilon_{1:k-1}} \mathds{P}_\gamma(\mathcal{G}^-)\theta(i,j | \mathcal{G}^-) }_{m_{ij}} = \sum^{k-1}_{j=1} p'_{jk}\pi_{jk} m_{ij} \ ^{cf. \ref{gim:eq:gim-mik}} \nonumber
\end{align*}

\section{Efficient computation of \gim}
\label{gim:app:matrix-computation}

We define two matrices.
First, the transfer matrix $T = [ p'_{ij}*\pi_{ij} ]$, where the element $p'_{ij}$ is the probability that tweet $v_j$ is a direct retweet of tweet $v_i$ (defined in \cref{gim:eq:pij}) and $\pi_{ij}$ is the proportion of capital transfered from $v_j$ to $v_i$;
Second, the influence accumulation matrix $M = [ m_{ij} ]$, with $m_{ij}$ defined in Eq.~\eqref{gim:eq:gim-mik} is the contribution of $v_j$ to the influence of $v_i$.
For each column $j$ of $M$, we compute the first $j-1$ elements by multiplying the sub-matrix $M_{[1..j-1, 1..j-1]}$ with the first $j-1$ elements on the $j^{th}$ column of matrix $T$, the $j$-th element is $\pi_{jj}$, and the remaining elements are $0$.
The computation of matrix $M$ finishes after $n$ steps, where $n$ is cascade size.

\section{Value-Allocation Scheme}
Value-allocation schemes over networks are an example of fair division games, concerned with how to allocate the value generated by a network of players among the players; for example, the allocation of advertisement revenue over a chain of marketers, or the allocation of utility revenue across a network of electricity infrastructure providers. 
Allocation scheme literature generally aims to understand the stability, efficiency, and fairness of network formation (usually in cooperative undirected networks). Our paper is concerned with the allocation of \socialcapital in a non-cooperative setting over an ad-hoc directed-acyclic diffusion graph.
The seminal work of \citet{jackson2003strategic}'s \emph{Connections} game defines the utility of a player as the sum of benefits to all other players decayed by the length of the path between them minus the cost of maintaining direct links for the player. 

 \end{appendices}


\begin{thebibliography}{73}
\ifx \bisbn   \undefined \def \bisbn  #1{ISBN #1}\fi
\ifx \binits  \undefined \def \binits#1{#1}\fi
\ifx \bauthor  \undefined \def \bauthor#1{#1}\fi
\ifx \batitle  \undefined \def \batitle#1{#1}\fi
\ifx \bjtitle  \undefined \def \bjtitle#1{#1}\fi
\ifx \bvolume  \undefined \def \bvolume#1{\textbf{#1}}\fi
\ifx \byear  \undefined \def \byear#1{#1}\fi
\ifx \bissue  \undefined \def \bissue#1{#1}\fi
\ifx \bfpage  \undefined \def \bfpage#1{#1}\fi
\ifx \blpage  \undefined \def \blpage #1{#1}\fi
\ifx \burl  \undefined \def \burl#1{\textsf{#1}}\fi
\ifx \doiurl  \undefined \def \doiurl#1{\url{https://doi.org/#1}}\fi
\ifx \betal  \undefined \def \betal{\textit{et al.}}\fi
\ifx \binstitute  \undefined \def \binstitute#1{#1}\fi
\ifx \binstitutionaled  \undefined \def \binstitutionaled#1{#1}\fi
\ifx \bctitle  \undefined \def \bctitle#1{#1}\fi
\ifx \beditor  \undefined \def \beditor#1{#1}\fi
\ifx \bpublisher  \undefined \def \bpublisher#1{#1}\fi
\ifx \bbtitle  \undefined \def \bbtitle#1{#1}\fi
\ifx \bedition  \undefined \def \bedition#1{#1}\fi
\ifx \bseriesno  \undefined \def \bseriesno#1{#1}\fi
\ifx \blocation  \undefined \def \blocation#1{#1}\fi
\ifx \bsertitle  \undefined \def \bsertitle#1{#1}\fi
\ifx \bsnm \undefined \def \bsnm#1{#1}\fi
\ifx \bsuffix \undefined \def \bsuffix#1{#1}\fi
\ifx \bparticle \undefined \def \bparticle#1{#1}\fi
\ifx \barticle \undefined \def \barticle#1{#1}\fi
\bibcommenthead
\ifx \bconfdate \undefined \def \bconfdate #1{#1}\fi
\ifx \botherref \undefined \def \botherref #1{#1}\fi
\ifx \url \undefined \def \url#1{\textsf{#1}}\fi
\ifx \bchapter \undefined \def \bchapter#1{#1}\fi
\ifx \bbook \undefined \def \bbook#1{#1}\fi
\ifx \bcomment \undefined \def \bcomment#1{#1}\fi
\ifx \oauthor \undefined \def \oauthor#1{#1}\fi
\ifx \citeauthoryear \undefined \def \citeauthoryear#1{#1}\fi
\ifx \endbibitem  \undefined \def \endbibitem {}\fi
\ifx \bconflocation  \undefined \def \bconflocation#1{#1}\fi
\ifx \arxivurl  \undefined \def \arxivurl#1{\textsf{#1}}\fi
\csname PreBibitemsHook\endcsname

\bibitem[\protect\citeauthoryear{Rapp}{2016}]{rapp2016moral}
\begin{barticle}
\bauthor{\bsnm{Rapp}, \binits{C.}}:
\batitle{Moral opinion polarization and the erosion of trust}.
\bjtitle{Social science research}
\bvolume{58},
\bfpage{34}--\blpage{45}
(\byear{2016})
\end{barticle}
\endbibitem

\bibitem[\protect\citeauthoryear{McCauley and
  Moskalenko}{2008}]{mccauley2008mechanisms}
\begin{barticle}
\bauthor{\bsnm{McCauley}, \binits{C.}},
\bauthor{\bsnm{Moskalenko}, \binits{S.}}:
\batitle{Mechanisms of political radicalization: Pathways toward terrorism}.
\bjtitle{Terrorism and political violence}
\bvolume{20}(\bissue{3}),
\bfpage{415}--\blpage{433}
(\byear{2008})
\end{barticle}
\endbibitem

\bibitem[\protect\citeauthoryear{Nowak et~al.}{2005}]{nowak2005dynamics}
\begin{botherref}
\oauthor{\bsnm{Nowak}, \binits{A.}},
\oauthor{\bsnm{Vallacher}, \binits{R.R.}},
\oauthor{\bsnm{Kus}, \binits{M.}},
\oauthor{\bsnm{Urbaniak}, \binits{J.}}:
The dynamics of societal transition: Modeling nonlinear change in the polish
  economic system.
International Journal of Sociology
(2005)
\end{botherref}
\endbibitem

\bibitem[\protect\citeauthoryear{Greijdanus
  et~al.}{2020}]{greijdanus2020psychology}
\begin{botherref}
\oauthor{\bsnm{Greijdanus}, \binits{H.}},
\oauthor{\bsnm{Matos~Fernandes}, \binits{C.A.}},
\oauthor{\bsnm{Turner-Zwinkels}, \binits{F.}},
\oauthor{\bsnm{Honari}, \binits{A.}},
\oauthor{\bsnm{Roos}, \binits{C.A.}},
\oauthor{\bsnm{Rosenbusch}, \binits{H.}},
\oauthor{\bsnm{Postmes}, \binits{T.}}:
The psychology of online activism and social movements: Relations between
  online and offline collective action.
Current opinion in psychology
(2020)
\end{botherref}
\endbibitem

\bibitem[\protect\citeauthoryear{Gunton}{2022}]{gunton2022impact}
\begin{bchapter}
\bauthor{\bsnm{Gunton}, \binits{K.}}:
\bctitle{The impact of the internet and social media platforms on
  radicalisation to terrorism and violent extremism}.
In: \bbtitle{Privacy, Security And Forensics in The Internet of Things (IoT)},
(\byear{2022})
\end{bchapter}
\endbibitem

\bibitem[\protect\citeauthoryear{Peng et~al.}{2018}]{peng2018influence}
\begin{botherref}
\oauthor{\bsnm{Peng}, \binits{S.}},
\oauthor{\bsnm{Zhou}, \binits{Y.}},
\oauthor{\bsnm{Cao}, \binits{L.}},
\oauthor{\bsnm{Yu}, \binits{S.}},
\oauthor{\bsnm{Niu}, \binits{J.}},
\oauthor{\bsnm{Jia}, \binits{W.}}:
Influence analysis in social networks: A survey.
Journal of Network and Computer Applications
(2018)
\end{botherref}
\endbibitem

\bibitem[\protect\citeauthoryear{Mason et~al.}{2007}]{mason2007situating}
\begin{botherref}
\oauthor{\bsnm{Mason}, \binits{W.A.}},
\oauthor{\bsnm{Conrey}, \binits{F.R.}},
\oauthor{\bsnm{Smith}, \binits{E.R.}}:
Situating social influence processes: Dynamic, multidirectional flows of
  influence within social networks.
Personality and social psychology review
(2007)
\end{botherref}
\endbibitem

\bibitem[\protect\citeauthoryear{Kempe et~al.}{2003}]{kempe2003maximizing}
\begin{bchapter}
\bauthor{\bsnm{Kempe}, \binits{D.}},
\bauthor{\bsnm{Kleinberg}, \binits{J.}},
\bauthor{\bsnm{Tardos}, \binits{{\'E}.}}:
\bctitle{Maximizing the spread of influence through a social network}.
In: \bbtitle{KDD}
(\byear{2003})
\end{bchapter}
\endbibitem

\bibitem[\protect\citeauthoryear{Du et~al.}{2013}]{du2013scalable}
\begin{bchapter}
\bauthor{\bsnm{Du}, \binits{N.}},
\bauthor{\bsnm{Song}, \binits{L.}},
\bauthor{\bsnm{Gomez-Rodriguez}, \binits{M.}},
\bauthor{\bsnm{Zha}, \binits{H.}}:
\bctitle{Scalable influence estimation in continuous-time diffusion networks}.
In: \bbtitle{NIPS}
(\byear{2013})
\end{bchapter}
\endbibitem

\bibitem[\protect\citeauthoryear{Mishra et~al.}{2016}]{mishra2016feature}
\begin{bchapter}
\bauthor{\bsnm{Mishra}, \binits{S.}},
\bauthor{\bsnm{Rizoiu}, \binits{M.-A.}},
\bauthor{\bsnm{Xie}, \binits{L.}}:
\bctitle{Feature driven and point process approaches for popularity
  prediction}.
In: \bbtitle{CIKM}
(\byear{2016})
\end{bchapter}
\endbibitem

\bibitem[\protect\citeauthoryear{Moussa{\"\i}d
  et~al.}{2013}]{moussaid2013social}
\begin{barticle}
\bauthor{\bsnm{Moussa{\"\i}d}, \binits{M.}},
\bauthor{\bsnm{K{\"a}mmer}, \binits{J.E.}},
\bauthor{\bsnm{Analytis}, \binits{P.P.}},
\bauthor{\bsnm{Neth}, \binits{H.}}:
\batitle{Social influence and the collective dynamics of opinion formation}.
\bjtitle{PloS one}
\bvolume{8}(\bissue{11}),
\bfpage{78433}
(\byear{2013})
\end{barticle}
\endbibitem

\bibitem[\protect\citeauthoryear{Kraus et~al.}{2019}]{kraus2019social}
\begin{botherref}
\oauthor{\bsnm{Kraus}, \binits{M.W.}},
\oauthor{\bsnm{Callaghan}, \binits{B.}},
\oauthor{\bsnm{Ondish}, \binits{P.}}:
Social class as culture.
(2019)
\end{botherref}
\endbibitem

\bibitem[\protect\citeauthoryear{Cialdini}{2001}]{cialdini2001science}
\begin{botherref}
\oauthor{\bsnm{Cialdini}, \binits{R.B.}}:
Influence: Science and practice.
Pearson
(2001)
\end{botherref}
\endbibitem

\bibitem[\protect\citeauthoryear{Asch}{1961}]{asch1961effects}
\begin{bchapter}
\bauthor{\bsnm{Asch}, \binits{S.E.}}:
\bctitle{Effects of group pressure upon the modification and distortion of
  judgments}.
In: \bbtitle{Documents of Gestalt Psychology},
(\byear{1961})
\end{bchapter}
\endbibitem

\bibitem[\protect\citeauthoryear{Rizoiu et~al.}{2018}]{rizoiu2018debatenight}
\begin{bchapter}
\bauthor{\bsnm{Rizoiu}, \binits{M.-A.}},
\bauthor{\bsnm{Graham}, \binits{T.}},
\bauthor{\bsnm{Zhang}, \binits{R.}},
\bauthor{\bsnm{Zhang}, \binits{Y.}},
\bauthor{\bsnm{Ackland}, \binits{R.}},
\bauthor{\bsnm{Xie}, \binits{L.}}:
\bctitle{\# debatenight: The role and influence of socialbots on twitter during
  the 1st 2016 us presidential debate}.
In: \bbtitle{ICWSM}
(\byear{2018})
\end{bchapter}
\endbibitem

\bibitem[\protect\citeauthoryear{Newcomb}{1953}]{newcomb1953approach}
\begin{botherref}
\oauthor{\bsnm{Newcomb}, \binits{T.M.}}:
An approach to the study of communicative acts.
Psychological review
(1953)
\end{botherref}
\endbibitem

\bibitem[\protect\citeauthoryear{Milgram and
  Gudehus}{1978}]{milgram1978obedience}
\begin{botherref}
\oauthor{\bsnm{Milgram}, \binits{S.}},
\oauthor{\bsnm{Gudehus}, \binits{C.}}:
Obedience to authority.
Ziff-Davis Publishing Company
(1978)
\end{botherref}
\endbibitem

\bibitem[\protect\citeauthoryear{Horai et~al.}{1974}]{horai1974effects}
\begin{botherref}
\oauthor{\bsnm{Horai}, \binits{J.}},
\oauthor{\bsnm{Naccari}, \binits{N.}},
\oauthor{\bsnm{Fatoullah}, \binits{E.}}:
The effects of expertise and physical attractiveness upon opinion agreement and
  liking.
Sociometry
(1974)
\end{botherref}
\endbibitem

\bibitem[\protect\citeauthoryear{Ghaffar and
  Hurley}{2020}]{ghaffar2020structural}
\begin{botherref}
\oauthor{\bsnm{Ghaffar}, \binits{F.}},
\oauthor{\bsnm{Hurley}, \binits{N.}}:
Structural hole centrality: evaluating social capital through strategic network
  formation.
Computational Social Networks
(2020)
\end{botherref}
\endbibitem

\bibitem[\protect\citeauthoryear{Sloan et~al.}{2015}]{sloan2015tweets}
\begin{botherref}
\oauthor{\bsnm{Sloan}, \binits{L.}},
\oauthor{\bsnm{Morgan}, \binits{J.}},
\oauthor{\bsnm{Burnap}, \binits{P.}},
\oauthor{\bsnm{Williams}, \binits{M.}}:
Who tweets? deriving the demographic characteristics of age, occupation and
  social class from twitter user meta-data.
PloS one
(2015)
\end{botherref}
\endbibitem

\bibitem[\protect\citeauthoryear{for O*NET~Development}{}]{onet}
\begin{botherref}
\oauthor{\bsnm{O*NET~Development}, \binits{N.C.}}:
O*NET OnLine,.
\url{https://www.onetonline.org/}.
Accessed: 2021-06-30
\end{botherref}
\endbibitem

\bibitem[\protect\citeauthoryear{Cui and Lee}{2020}]{cui2020coaid}
\begin{botherref}
\oauthor{\bsnm{Cui}, \binits{L.}},
\oauthor{\bsnm{Lee}, \binits{D.}}:
Coaid: Covid-19 healthcare misinformation dataset.
arXiv preprint arXiv:2006.00885
(2020)
\end{botherref}
\endbibitem

\bibitem[\protect\citeauthoryear{Metaxas et~al.}{2015}]{metaxas2015retweets}
\begin{bchapter}
\bauthor{\bsnm{Metaxas}, \binits{P.}},
\bauthor{\bsnm{Mustafaraj}, \binits{E.}},
\bauthor{\bsnm{Wong}, \binits{K.}},
\bauthor{\bsnm{Zeng}, \binits{L.}},
\bauthor{\bsnm{O'Keefe}, \binits{M.}},
\bauthor{\bsnm{Finn}, \binits{S.}}:
\bctitle{What do retweets indicate? results from user survey and meta-review of
  research}.
In: \bbtitle{Proceedings of the International AAAI Conference on Web and Social
  Media},
vol. \bseriesno{9},
pp. \bfpage{658}--\blpage{661}
(\byear{2015})
\end{bchapter}
\endbibitem

\bibitem[\protect\citeauthoryear{Liu et~al.}{2017}]{liu2017influence}
\begin{botherref}
\oauthor{\bsnm{Liu}, \binits{Q.}},
\oauthor{\bsnm{Xiang}, \binits{B.}},
\oauthor{\bsnm{Yuan}, \binits{N.J.}},
\oauthor{\bsnm{Chen}, \binits{E.}},
\oauthor{\bsnm{Xiong}, \binits{H.}},
\oauthor{\bsnm{Zheng}, \binits{Y.}},
\oauthor{\bsnm{Yang}, \binits{Y.}}:
An influence propagation view of pagerank.
TKDD
(2017)
\end{botherref}
\endbibitem

\bibitem[\protect\citeauthoryear{Silva et~al.}{2013}]{silva2013profilerank}
\begin{bchapter}
\bauthor{\bsnm{Silva}, \binits{A.}},
\bauthor{\bsnm{Guimar{\~a}es}, \binits{S.}},
\bauthor{\bsnm{Meira~Jr}, \binits{W.}},
\bauthor{\bsnm{Zaki}, \binits{M.}}:
\bctitle{Profilerank: finding relevant content and influential users based on
  information diffusion}.
In: \bbtitle{SNAKDD}
(\byear{2013})
\end{bchapter}
\endbibitem

\bibitem[\protect\citeauthoryear{Mishra et~al.}{2018}]{mishra2018modeling}
\begin{bchapter}
\bauthor{\bsnm{Mishra}, \binits{S.}},
\bauthor{\bsnm{Rizoiu}, \binits{M.-A.}},
\bauthor{\bsnm{Xie}, \binits{L.}}:
\bctitle{Modeling popularity in asynchronous social media streams with
  recurrent neural networks}.
In: \bbtitle{ICWSM}
(\byear{2018})
\end{bchapter}
\endbibitem

\bibitem[\protect\citeauthoryear{Romero et~al.}{2011}]{romero2011influence}
\begin{bchapter}
\bauthor{\bsnm{Romero}, \binits{D.M.}},
\bauthor{\bsnm{Galuba}, \binits{W.}},
\bauthor{\bsnm{Asur}, \binits{S.}},
\bauthor{\bsnm{Huberman}, \binits{B.A.}}:
\bctitle{Influence and passivity in social media}.
In: \bbtitle{ECML PKDD}
(\byear{2011})
\end{bchapter}
\endbibitem

\bibitem[\protect\citeauthoryear{Bakshy et~al.}{2011}]{bakshy2011everyone}
\begin{bchapter}
\bauthor{\bsnm{Bakshy}, \binits{E.}},
\bauthor{\bsnm{Hofman}, \binits{J.M.}},
\bauthor{\bsnm{Mason}, \binits{W.A.}},
\bauthor{\bsnm{Watts}, \binits{D.J.}}:
\bctitle{Everyone's an influencer: quantifying influence on twitter}.
In: \bbtitle{WSDM}
(\byear{2011})
\end{bchapter}
\endbibitem

\bibitem[\protect\citeauthoryear{Smith et~al.}{2018}]{smith2018influence}
\begin{bchapter}
\bauthor{\bsnm{Smith}, \binits{S.T.}},
\bauthor{\bsnm{Kao}, \binits{E.K.}},
\bauthor{\bsnm{Shah}, \binits{D.C.}},
\bauthor{\bsnm{Simek}, \binits{O.}},
\bauthor{\bsnm{Rubin}, \binits{D.B.}}:
\bctitle{Influence estimation on social media networks using causal inference}.
In: \bbtitle{IEEE SSP}
(\byear{2018})
\end{bchapter}
\endbibitem

\bibitem[\protect\citeauthoryear{Nickel and Le}{2021}]{nickel2021modeling}
\begin{bchapter}
\bauthor{\bsnm{Nickel}, \binits{M.}},
\bauthor{\bsnm{Le}, \binits{M.}}:
\bctitle{Modeling sparse information diffusion at scale via lazy multivariate
  hawkes processes}.
In: \bbtitle{WWW}
(\byear{2021})
\end{bchapter}
\endbibitem

\bibitem[\protect\citeauthoryear{Hawkes}{1971}]{hawkes1971spectra}
\begin{botherref}
\oauthor{\bsnm{Hawkes}, \binits{A.G.}}:
Spectra of some self-exciting and mutually exciting point processes.
Biometrika
(1971)
\end{botherref}
\endbibitem

\bibitem[\protect\citeauthoryear{Rizoiu et~al.}{2017}]{Rizoiu2017a}
\begin{bbook}
\bauthor{\bsnm{Rizoiu}, \binits{M.-A.}},
\bauthor{\bsnm{Lee}, \binits{Y.}},
\bauthor{\bsnm{Mishra}, \binits{S.}},
\bauthor{\bsnm{Xie}, \binits{L.}}:
In: \beditor{\bsnm{Chang}, \binits{S.-F.}} (ed.)
\bbtitle{Hawkes processes for events in social media},
pp. \bfpage{191}--\blpage{218}.
\bpublisher{Association for Computing Machinery and Morgan \& Claypool},
  \blocation{???}
(\byear{2017}).
\doiurl{10.1145/3122865.3122874} .
\burl{https://dl.acm.org/doi/10.1145/3122865.3122874}
\end{bbook}
\endbibitem

\bibitem[\protect\citeauthoryear{Kong et~al.}{2020}]{Kong2020}
\begin{bchapter}
\bauthor{\bsnm{Kong}, \binits{Q.}},
\bauthor{\bsnm{Rizoiu}, \binits{M.-A.}},
\bauthor{\bsnm{Xie}, \binits{L.}}:
\bctitle{Modeling information cascades with self-exciting processes via
  generalized epidemic models}.
In: \bbtitle{Proceedings of the 13th International Conference on Web Search and
  Data Mining},
pp. \bfpage{286}--\blpage{294}.
\bpublisher{ACM}, \blocation{???}
(\byear{2020}).
\doiurl{10.1145/3336191.3371821} .
\burl{https://arxiv.org/abs/1910.05451
  https://dl.acm.org/doi/10.1145/3336191.3371821}
\end{bchapter}
\endbibitem

\bibitem[\protect\citeauthoryear{Hawkes and Oakes}{1974}]{hawkes1974cluster}
\begin{botherref}
\oauthor{\bsnm{Hawkes}, \binits{A.G.}},
\oauthor{\bsnm{Oakes}, \binits{D.}}:
A cluster process representation of a self-exciting process.
Journal of Applied Probability
(1974)
\end{botherref}
\endbibitem

\bibitem[\protect\citeauthoryear{Lewis and
  Mohler}{2011}]{lewis2011nonparametric}
\begin{botherref}
\oauthor{\bsnm{Lewis}, \binits{E.}},
\oauthor{\bsnm{Mohler}, \binits{G.}}:
A nonparametric em algorithm for multiscale hawkes processes.
Journal of Nonparametric Statistics
(2011)
\end{botherref}
\endbibitem

\bibitem[\protect\citeauthoryear{Luceri et~al.}{2019}]{luceri2019analyzing}
\begin{botherref}
\oauthor{\bsnm{Luceri}, \binits{L.}},
\oauthor{\bsnm{Braun}, \binits{T.}},
\oauthor{\bsnm{Giordano}, \binits{S.}}:
Analyzing and inferring human real-life behavior through online social networks
  with social influence deep learning.
Applied network science
(2019)
\end{botherref}
\endbibitem

\bibitem[\protect\citeauthoryear{Qiu et~al.}{2018}]{qiu2018deepinf}
\begin{bchapter}
\bauthor{\bsnm{Qiu}, \binits{J.}},
\bauthor{\bsnm{Tang}, \binits{J.}},
\bauthor{\bsnm{Ma}, \binits{H.}},
\bauthor{\bsnm{Dong}, \binits{Y.}},
\bauthor{\bsnm{Wang}, \binits{K.}},
\bauthor{\bsnm{Tang}, \binits{J.}}:
\bctitle{Deepinf: Social influence prediction with deep learning}.
In: \bbtitle{KDD}
(\byear{2018})
\end{bchapter}
\endbibitem

\bibitem[\protect\citeauthoryear{Gu et~al.}{2018}]{gu2018rare}
\begin{bchapter}
\bauthor{\bsnm{Gu}, \binits{Y.}},
\bauthor{\bsnm{Sun}, \binits{Y.}},
\bauthor{\bsnm{Li}, \binits{Y.}},
\bauthor{\bsnm{Yang}, \binits{Y.}}:
\bctitle{Rare: Social rank regulated large-scale network embedding}.
In: \bbtitle{WWW}
(\byear{2018})
\end{bchapter}
\endbibitem

\bibitem[\protect\citeauthoryear{Tsitsulin et~al.}{2018}]{tsitsulin2018verse}
\begin{bchapter}
\bauthor{\bsnm{Tsitsulin}, \binits{A.}},
\bauthor{\bsnm{Mottin}, \binits{D.}},
\bauthor{\bsnm{Karras}, \binits{P.}},
\bauthor{\bsnm{M{\"u}ller}, \binits{E.}}:
\bctitle{Verse: Versatile graph embeddings from similarity measures}.
In: \bbtitle{WWW}
(\byear{2018})
\end{bchapter}
\endbibitem

\bibitem[\protect\citeauthoryear{Yu and Qin}{2019}]{yu2019spectrum}
\begin{bchapter}
\bauthor{\bsnm{Yu}, \binits{W.}},
\bauthor{\bsnm{Qin}, \binits{Z.}}:
\bctitle{Spectrum-enhanced pairwise learning to rank}.
In: \bbtitle{WWW}
(\byear{2019})
\end{bchapter}
\endbibitem

\bibitem[\protect\citeauthoryear{Rizoiu et~al.}{2018}]{Rizoiu2018}
\begin{bchapter}
\bauthor{\bsnm{Rizoiu}, \binits{M.-A.}},
\bauthor{\bsnm{Mishra}, \binits{S.}},
\bauthor{\bsnm{Kong}, \binits{Q.}},
\bauthor{\bsnm{Carman}, \binits{M.}},
\bauthor{\bsnm{Xie}, \binits{L.}}:
\bctitle{Sir-hawkes: Linking epidemic models and hawkes processes to model
  diffusions in finite populations}.
In: \bbtitle{Proceedings of the 2018 World Wide Web Conference on World Wide
  Web - WWW '18},
pp. \bfpage{419}--\blpage{428}.
\bpublisher{ACM Press}, \blocation{???}
(\byear{2018}).
\doiurl{10.1145/3178876.3186108} .
\burl{http://arxiv.org/abs/1711.01679
  http://dl.acm.org/citation.cfm?doid=3178876.3186108}
\end{bchapter}
\endbibitem

\bibitem[\protect\citeauthoryear{Lin et~al.}{2018}]{lin2018intergroup}
\begin{botherref}
\oauthor{\bsnm{Lin}, \binits{L.C.}},
\oauthor{\bsnm{Qu}, \binits{Y.}},
\oauthor{\bsnm{Telzer}, \binits{E.H.}}:
Intergroup social influence on emotion processing in the brain.
PNAS
(2018)
\end{botherref}
\endbibitem

\bibitem[\protect\citeauthoryear{Centola}{2011}]{centola2011experimental}
\begin{barticle}
\bauthor{\bsnm{Centola}, \binits{D.}}:
\batitle{An experimental study of homophily in the adoption of health
  behavior}.
\bjtitle{Science}
\bvolume{334}(\bissue{6060}),
\bfpage{1269}--\blpage{1272}
(\byear{2011})
\end{barticle}
\endbibitem

\bibitem[\protect\citeauthoryear{Crandall et~al.}{2008}]{crandall2008feedback}
\begin{bchapter}
\bauthor{\bsnm{Crandall}, \binits{D.}},
\bauthor{\bsnm{Cosley}, \binits{D.}},
\bauthor{\bsnm{Huttenlocher}, \binits{D.}},
\bauthor{\bsnm{Kleinberg}, \binits{J.}},
\bauthor{\bsnm{Suri}, \binits{S.}}:
\bctitle{Feedback effects between similarity and social influence in online
  communities}.
In: \bbtitle{KDD}
(\byear{2008})
\end{bchapter}
\endbibitem

\bibitem[\protect\citeauthoryear{Goel et~al.}{2016}]{goel2016social}
\begin{bchapter}
\bauthor{\bsnm{Goel}, \binits{R.}},
\bauthor{\bsnm{Soni}, \binits{S.}},
\bauthor{\bsnm{Goyal}, \binits{N.}},
\bauthor{\bsnm{Paparrizos}, \binits{J.}},
\bauthor{\bsnm{Wallach}, \binits{H.}},
\bauthor{\bsnm{Diaz}, \binits{F.}},
\bauthor{\bsnm{Eisenstein}, \binits{J.}}:
\bctitle{The social dynamics of language change in online networks}.
In: \bbtitle{International Conference on Social Informatics}
(\byear{2016})
\end{bchapter}
\endbibitem

\bibitem[\protect\citeauthoryear{Moody}{1989}]{moody1989fast}
\begin{bchapter}
\bauthor{\bsnm{Moody}, \binits{J.}}:
\bctitle{Fast learning in multi-resolution hierarchies}.
In: \bbtitle{NIPS}
(\byear{1989})
\end{bchapter}
\endbibitem

\bibitem[\protect\citeauthoryear{Granovetter}{1983}]{granovetter1983strength}
\begin{botherref}
\oauthor{\bsnm{Granovetter}, \binits{M.}}:
The strength of weak ties: A network theory revisited.
Sociological theory
(1983)
\end{botherref}
\endbibitem

\bibitem[\protect\citeauthoryear{Dekker and Uslaner}{2003}]{dekker2003social}
\begin{bbook}
\bauthor{\bsnm{Dekker}, \binits{P.}},
\bauthor{\bsnm{Uslaner}, \binits{E.M.}}:
\bbtitle{Social Capital and Participation in Everyday Life}.
\bpublisher{Routledge},
\blocation{Oxfordshire}
(\byear{2003})
\end{bbook}
\endbibitem

\bibitem[\protect\citeauthoryear{Shapley}{1997}]{shapley1997value}
\begin{botherref}
\oauthor{\bsnm{Shapley}, \binits{L.S.}}:
A value for n-person games.
Classics in game theory
\textbf{69}
(1997)
\end{botherref}
\endbibitem

\bibitem[\protect\citeauthoryear{Jackson and
  Wolinsky}{2003}]{jackson2003strategic}
\begin{bchapter}
\bauthor{\bsnm{Jackson}, \binits{M.O.}},
\bauthor{\bsnm{Wolinsky}, \binits{A.}}:
\bctitle{A strategic model of social and economic networks}.
In: \bbtitle{Networks and Groups},
pp. \bfpage{23}--\blpage{49}.
\bpublisher{Springer},
\blocation{New York}
(\byear{2003})
\end{bchapter}
\endbibitem

\bibitem[\protect\citeauthoryear{Subbian et~al.}{2014}]{subbian2014finding}
\begin{botherref}
\oauthor{\bsnm{Subbian}, \binits{K.}},
\oauthor{\bsnm{Sharma}, \binits{D.}},
\oauthor{\bsnm{Wen}, \binits{Z.}},
\oauthor{\bsnm{Srivastava}, \binits{J.}}:
Finding influencers in networks using social capital.
Soc. Net. Analysis \& Mining
(2014)
\end{botherref}
\endbibitem

\bibitem[\protect\citeauthoryear{Appendix}{2021}]{appendix}
\begin{botherref}
\oauthor{\bsnm{Appendix}, \binits{O.}}:
Appendix: \titlename.
\url{https://www.dropbox.com/s/wcaovmb5oeawnvq/supplementary_material.pdf?dl=0}
(2021)
\end{botherref}
\endbibitem

\bibitem[\protect\citeauthoryear{Ram and Rizoiu}{2024}]{empiricalMeasurement}
\begin{barticle}
\bauthor{\bsnm{Ram}, \binits{R.}},
\bauthor{\bsnm{Rizoiu}, \binits{M.-A.}}:
\batitle{Empirically measuring online social influence}.
\bjtitle{EPJ Data Science}
\bvolume{13},
\bfpage{53}
(\byear{2024})
\doiurl{10.1140/epjds/s13688-024-00492-z}
\end{barticle}
\endbibitem

\bibitem[\protect\citeauthoryear{Carpentras and
  Quayle}{2023}]{carpentras2023psychometric}
\begin{barticle}
\bauthor{\bsnm{Carpentras}, \binits{D.}},
\bauthor{\bsnm{Quayle}, \binits{M.}}:
\batitle{The psychometric house-of-mirrors: the effect of measurement
  distortions on agent-based models' predictions}.
\bjtitle{International Journal of Social Research Methodology}
\bvolume{26}(\bissue{2}),
\bfpage{215}--\blpage{231}
(\byear{2023})
\end{barticle}
\endbibitem

\bibitem[\protect\citeauthoryear{Graham and Keller}{2020}]{graham2020bushfires}
\begin{botherref}
\oauthor{\bsnm{Graham}, \binits{T.}},
\oauthor{\bsnm{Keller}, \binits{T.}}:
Bushfires, bots and arson claims: Australia flung in the global disinformation
  spotlight.
The Conversation
\textbf{10}
(2020)
\end{botherref}
\endbibitem

\bibitem[\protect\citeauthoryear{Page et~al.}{1999}]{page1999pagerank}
\begin{botherref}
\oauthor{\bsnm{Page}, \binits{L.}},
\oauthor{\bsnm{Brin}, \binits{S.}},
\oauthor{\bsnm{Motwani}, \binits{R.}},
\oauthor{\bsnm{Winograd}, \binits{T.}}:
The pagerank citation ranking: Bringing order to the web.
Technical report,
Stanford
(1999)
\end{botherref}
\endbibitem

\bibitem[\protect\citeauthoryear{Xiang et~al.}{2013}]{xiang2013pagerank}
\begin{bchapter}
\bauthor{\bsnm{Xiang}, \binits{B.}},
\bauthor{\bsnm{Liu}, \binits{Q.}},
\bauthor{\bsnm{Chen}, \binits{E.}},
\bauthor{\bsnm{Xiong}, \binits{H.}},
\bauthor{\bsnm{Zheng}, \binits{Y.}},
\bauthor{\bsnm{Yang}, \binits{Y.}}:
\bctitle{Pagerank with priors: An influence propagation perspective}.
In: \bbtitle{IJCAI}
(\byear{2013})
\end{bchapter}
\endbibitem

\bibitem[\protect\citeauthoryear{Cha et~al.}{2010}]{cha2010measuring}
\begin{bchapter}
\bauthor{\bsnm{Cha}, \binits{M.}},
\bauthor{\bsnm{Haddadi}, \binits{H.}},
\bauthor{\bsnm{Benevenuto}, \binits{F.}},
\bauthor{\bsnm{Gummadi}, \binits{K.}}:
\bctitle{Measuring user influence in twitter: The million follower fallacy}.
In: \bbtitle{ICWSM}
(\byear{2010})
\end{bchapter}
\endbibitem

\bibitem[\protect\citeauthoryear{Wu et~al.}{2019}]{Wu2019}
\begin{barticle}
\bauthor{\bsnm{Wu}, \binits{S.}},
\bauthor{\bsnm{Rizoiu}, \binits{M.-A.}},
\bauthor{\bsnm{Xie}, \binits{L.}}:
\batitle{Estimating attention flow in online video networks}.
\bjtitle{Proceedings of the ACM on Human-Computer Interaction}
\bvolume{3},
\bfpage{1}--\blpage{25}
(\byear{2019})
\doiurl{10.1145/3359285}
\end{barticle}
\endbibitem

\bibitem[\protect\citeauthoryear{Frantz et~al.}{2009}]{frantz2009robustness}
\begin{barticle}
\bauthor{\bsnm{Frantz}, \binits{T.L.}},
\bauthor{\bsnm{Cataldo}, \binits{M.}},
\bauthor{\bsnm{Carley}, \binits{K.M.}}:
\batitle{Robustness of centrality measures under uncertainty: Examining the
  role of network topology}.
\bjtitle{Computational and Mathematical Organization Theory}
\bvolume{15}(\bissue{4}),
\bfpage{303}--\blpage{328}
(\byear{2009})
\end{barticle}
\endbibitem

\bibitem[\protect\citeauthoryear{Riddell et~al.}{2017}]{riddell2017most}
\begin{botherref}
\oauthor{\bsnm{Riddell}, \binits{J.}},
\oauthor{\bsnm{Brown}, \binits{A.}},
\oauthor{\bsnm{Kovic}, \binits{I.}},
\oauthor{\bsnm{Jauregui}, \binits{J.}}:
Who are the most influential emergency physicians on twitter?
Western Journal of Emergency Medicine
(2017)
\end{botherref}
\endbibitem

\bibitem[\protect\citeauthoryear{Bailo et~al.}{2023}]{Bailo2023}
\begin{botherref}
\oauthor{\bsnm{Bailo}, \binits{F.}},
\oauthor{\bsnm{Johns}, \binits{A.}},
\oauthor{\bsnm{Rizoiu}, \binits{M.-A.}}:
Riding information crises: the performance of far-right twitter users in
  australia during the 2019–2020 bushfires and the covid-19 pandemic.
Information, Communication \& Society,
1--19
(2023)
\doiurl{10.1080/1369118X.2023.2205479}
\end{botherref}
\endbibitem

\bibitem[\protect\citeauthoryear{Kern et~al.}{2019}]{Kern2019}
\begin{barticle}
\bauthor{\bsnm{Kern}, \binits{M.L.}},
\bauthor{\bsnm{McCarthy}, \binits{P.X.}},
\bauthor{\bsnm{Chakrabarty}, \binits{D.}},
\bauthor{\bsnm{Rizoiu}, \binits{M.-A.}}:
\batitle{Social media-predicted personality traits and values can help match
  people to their ideal jobs}.
\bjtitle{Proceedings of the National Academy of Sciences}
\bvolume{116},
\bfpage{26459}--\blpage{26464}
(\byear{2019})
\doiurl{10.1073/pnas.1917942116}
\end{barticle}
\endbibitem

\bibitem[\protect\citeauthoryear{Turrell et~al.}{2018}]{turrell2018using}
\begin{botherref}
\oauthor{\bsnm{Turrell}, \binits{A.}},
\oauthor{\bsnm{Speigner}, \binits{B.}},
\oauthor{\bsnm{Djumalieva}, \binits{J.}},
\oauthor{\bsnm{Copple}, \binits{D.}},
\oauthor{\bsnm{Thurgood}, \binits{J.}}:
Using job vacancies to understand the effects of labour market mismatch on uk
  output and productivity
(2018)
\end{botherref}
\endbibitem

\bibitem[\protect\citeauthoryear{Mukherjee
  et~al.}{2021}]{mukherjee2021determining}
\begin{bchapter}
\bauthor{\bsnm{Mukherjee}, \binits{S.}},
\bauthor{\bsnm{Widmark}, \binits{D.}},
\bauthor{\bsnm{DiMascio}, \binits{V.}},
\bauthor{\bsnm{Oates}, \binits{T.}}:
\bctitle{Determining standard occupational classification codes from job
  descriptions in immigration petitions}.
In: \bbtitle{ICDMW}
(\byear{2021})
\end{bchapter}
\endbibitem

\bibitem[\protect\citeauthoryear{Singh et~al.}{2020}]{singh2020understanding}
\begin{botherref}
\oauthor{\bsnm{Singh}, \binits{L.}},
\oauthor{\bsnm{Bode}, \binits{L.}},
\oauthor{\bsnm{Budak}, \binits{C.}},
\oauthor{\bsnm{Kawintiranon}, \binits{K.}},
\oauthor{\bsnm{Padden}, \binits{C.}},
\oauthor{\bsnm{Vraga}, \binits{E.}}:
Understanding high-and low-quality url sharing on covid-19 twitter streams.
Journal of computational social science
(2020)
\end{botherref}
\endbibitem

\bibitem[\protect\citeauthoryear{Aral et~al.}{2009}]{aral2009distinguishing}
\begin{barticle}
\bauthor{\bsnm{Aral}, \binits{S.}},
\bauthor{\bsnm{Muchnik}, \binits{L.}},
\bauthor{\bsnm{Sundararajan}, \binits{A.}}:
\batitle{Distinguishing influence-based contagion from homophily-driven
  diffusion in dynamic networks}.
\bjtitle{Proceedings of the National Academy of Sciences}
\bvolume{106}(\bissue{51}),
\bfpage{21544}--\blpage{21549}
(\byear{2009})
\end{barticle}
\endbibitem

\bibitem[\protect\citeauthoryear{Romero et~al.}{2011}]{Romero2011}
\begin{bchapter}
\bauthor{\bsnm{Romero}, \binits{D.M.}},
\bauthor{\bsnm{Meeder}, \binits{B.}},
\bauthor{\bsnm{Kleinberg}, \binits{J.}}:
\bctitle{Differences in the mechanics of information diffusion across topics:
  idioms, political hashtags, and complex contagion on twitter}.
In: \bbtitle{Proceedings of the 20th International Conference on World Wide
  Web},
pp. \bfpage{695}--\blpage{704}
(\byear{2011}).
\doiurl{10.1145/1963405.1963503} .
\bcomment{ACM}
\end{bchapter}
\endbibitem

\bibitem[\protect\citeauthoryear{Bakshy et~al.}{2009}]{Bakshy2009}
\begin{bchapter}
\bauthor{\bsnm{Bakshy}, \binits{E.}},
\bauthor{\bsnm{Karrer}, \binits{B.}},
\bauthor{\bsnm{Adamic}, \binits{L.A.}}:
\bctitle{Social influence and the diffusion of user-created content}.
In: \bbtitle{Proceedings of the 10th ACM Conference on Electronic Commerce},
pp. \bfpage{325}--\blpage{334}
(\byear{2009}).
\doiurl{10.1145/1566374.1566421} .
\bcomment{ACM}
\end{bchapter}
\endbibitem

\bibitem[\protect\citeauthoryear{M{\o}nsted et~al.}{2017}]{Monsted2017}
\begin{barticle}
\bauthor{\bsnm{M{\o}nsted}, \binits{B.M.}},
\bauthor{\bsnm{Sapie{\.z}y{\'n}ski}, \binits{P.}},
\bauthor{\bsnm{Ferrara}, \binits{E.}},
\bauthor{\bsnm{Lehmann}, \binits{S.}}:
\batitle{Evidence of complex contagion of information in social media: An
  experiment using twitter bots}.
\bjtitle{PLOS ONE}
\bvolume{12}(\bissue{9}),
\bfpage{0184148}
(\byear{2017})
\doiurl{10.1371/journal.pone.0184148}
\end{barticle}
\endbibitem

\bibitem[\protect\citeauthoryear{Guilbeault et~al.}{2018}]{Guilbeault2018}
\begin{bchapter}
\bauthor{\bsnm{Guilbeault}, \binits{D.}},
\bauthor{\bsnm{Becker}, \binits{J.}},
\bauthor{\bsnm{Centola}, \binits{D.}}:
\bctitle{Complex contagions: A decade in review}.
In: \beditor{\bsnm{Lehmann}, \binits{S.}},
\beditor{\bsnm{Ahn}, \binits{Y.-Y.}} (eds.)
\bbtitle{Complex Spreading Phenomena in Social Systems},
pp. \bfpage{3}--\blpage{25}.
\bpublisher{Springer}, \blocation{???}
(\byear{2018}).
\doiurl{10.1007/978-3-319-77332-2_1}
\end{bchapter}
\endbibitem

\bibitem[\protect\citeauthoryear{Guilbeault and Centola}{2021}]{Guilbeault2021}
\begin{barticle}
\bauthor{\bsnm{Guilbeault}, \binits{D.}},
\bauthor{\bsnm{Centola}, \binits{D.}}:
\batitle{Topological measures for identifying and predicting the spread of
  complex contagions}.
\bjtitle{Nature Communications}
\bvolume{12},
\bfpage{4430}
(\byear{2021})
\doiurl{10.1038/s41467-021-24704-6}
\end{barticle}
\endbibitem

\bibitem[\protect\citeauthoryear{Kong et~al.}{2022}]{Kong2022}
\begin{barticle}
\bauthor{\bsnm{Kong}, \binits{Q.}},
\bauthor{\bsnm{Booth}, \binits{E.}},
\bauthor{\bsnm{Bailo}, \binits{F.}},
\bauthor{\bsnm{Johns}, \binits{A.}},
\bauthor{\bsnm{Rizoiu}, \binits{M.-A.}}:
\batitle{Slipping to the extreme: A mixed method to explain how extreme
  opinions infiltrate online discussions}.
\bjtitle{Proceedings of the International AAAI Conference on Web and Social
  Media}
\bvolume{16},
\bfpage{524}--\blpage{535}
(\byear{2022})
\doiurl{10.1609/icwsm.v16i1.19312}
\end{barticle}
\endbibitem

\end{thebibliography}
\end{document}